\font\tenmib=cmmib10
\font\sevenmib=cmmib10 scaled 800
\font\titolo=cmbx12
\font\titolone=cmbx10 scaled\magstep 2
\font\cs=cmcsc10

\font\ninerm=cmr9
\font\ottorm=cmr8
\textfont5=\tenmib
\scriptfont5=\sevenmib
\scriptscriptfont5=\fivei

\font\indbf=cmbx10 scaled\magstep2

\newcount\mgnf\newcount\driver
\driver=1 
\mgnf=0   

\ifnum\mgnf=0
   \magnification=\magstep0
   \hsize=14truecm\vsize=24.truecm
   \parindent=0.3cm\baselineskip=0.45cm\fi
\ifnum\mgnf=1
   \magnification=\magstep1\hoffset=0.truecm
   \hsize=14truecm\vsize=24.truecm
   \baselineskip=18truept plus0.1pt minus0.1pt \parindent=0.9truecm
   \lineskip=0.5truecm\lineskiplimit=0.1pt      \parskip=0.1pt plus1pt\fi

\let\a=\alpha \let\b=\beta  \let\g=\gamma     \let\d=\delta  \let\e=\varepsilon
\let\z=\zeta  \let\h=\eta    \let\k=\kappa   \let\l=\lambda
        \let\x=\xi        \let\p=\pi      
\let\s=\sigma \let\t=\tau   \let\f=\varphi    \let\ph=\varphi \let\c=\chi
               \let\o=\omega

 \let\D=\Delta     \let\L=\Lambda

\global\newcount\numsec\global\newcount\numfor

\def\SIA #1,#2,#3 {\senondefinito{#1#2}%
\expandafter\xdef\csname #1#2\endcsname{#3}\else
\write16{???? ma #1,#2 e' gia' stato definito !!!!} \fi}

\def \FU(#1)#2{\SIA fu,#1,#2 }

\def\etichetta(#1){(\veroparagrafo.\veraformula)%
\SIA e,#1,(\veroparagrafo.\veraformula) %
\global\advance\numfor by 1%
\write15{\string\FU (#1){\equ(#1)}}%
\write16{ EQ #1 ==> \equ(#1)  }}
\def\etichettaa(#1){(A\veraappendice.\veraformula)
 \SIA e,#1,(A\veraappendice.\veraformula)
 \global\advance\numfor by 1
 \write15{\string\FU (#1){\equ(#1)}}
 \write16{ EQ #1 ==> \equ(#1) }}
\def\getichetta(#1){Fig. \verafigura
 \SIA g,#1,{\verafigura}
 \global\advance\numfig by 1
 \write15{\string\FU (#1){\graf(#1)}}
 \write16{ Fig. #1 ==> \graf(#1) }}
\def\retichetta(#1){\numpag=\pgn\SIA r,#1,{\verapagina}
 \write15{\string\FU (#1){\rif(#1)}}
 \write16{\rif(#1) ha simbolo  #1  }}
\def\etichettan(#1){(n\verocapitolo.\veranformula)
 \SIA e,#1,(n\verocapitolo.\veranformula)
 \global\advance\numnf by 1
\write16{\equ(#1) <= #1  }}

\newdimen\gwidth
\gdef\profonditastruttura{\dp\strutbox}
\def\senondefinito#1{\expandafter\ifx\csname#1\endcsname\relax}
\def\BOZZA{
\def\alato(##1){
 {\vtop to \profonditastruttura{\baselineskip
 \profonditastruttura\vss
 \rlap{\kern-\hsize\kern-1.2truecm{$\scriptstyle##1$}}}}}
\def\galato(##1){ \gwidth=\hsize \divide\gwidth by 2
 {\vtop to \profonditastruttura{\baselineskip
 \profonditastruttura\vss
 \rlap{\kern-\gwidth\kern-1.2truecm{$\scriptstyle##1$}}}}}
\def\verapagina{
{\romannumeral\number\numcap}.\number\numsec.\number\numpag}}

\def\alato(#1){}
\def\galato(#1){}
\def\veroparagrafo{\number\numsec}\def\veraformula{\number\numfor}
\def\veraappendice{\number\numsec}
\def\verapagina{\number\pageno}\def\veranformula{\number\numnf}
\def\verafigura{{\romannumeral\number\numcap}.\number\numfig}
\def\verocapitolo{\number\numcap}\def\veranformula{\number\numnf}
\def\Eqn(#1){\eqno{\etichettan(#1)\alato(#1)}}
\def\eqn(#1){\etichettan(#1)\alato(#1)}

\def\Eq(#1){\eqno{\etichetta(#1)\alato(#1)}}
\def\eq(#1){\etichetta(#1)\alato(#1)}
\def\Eqa(#1){\eqno{\etichettaa(#1)\alato(#1)}}
\def\eqa(#1){\etichettaa(#1)\alato(#1)}

\def\eqv(#1){\senondefinito{fu#1}$\clubsuit$#1%
\write16{#1 non e' (ancora) definito}%
\else\csname fu#1\endcsname\fi}
\def\equ(#1){\senondefinito{e#1}\eqv(#1)\else\csname e#1\endcsname\fi}




\newcount\pgn \pgn=1
\def\foglio{\number\numsec:\number\pgn
\global\advance\pgn by 1}
\def\foglioa{A\number\numsec:\number\pgn
\global\advance\pgn by 1}

\footline={\rlap{\hbox{\copy200}}}

\newdimen\xshift \newdimen\xwidth
\def\ins#1#2#3{\vbox to0pt{\kern-#2 \hbox{\kern#1 #3}\vss}\nointerlineskip}
\def\insertplot#1#2#3#4#5{\par%
\xwidth=#1 \xshift=\hsize \advance\xshift by-\xwidth \divide\xshift by 2%
\yshift=#2 \divide\yshift by 2%
\line{\hskip\xshift \vbox to #2{\vfil%
\ifnum\driver=0 #3
\special{ps: plotfile #4.ps} 
\fi
\ifnum\driver=1 #3 \includegraphics{#4.ps}\fi
\ifnum\driver=2 #3
\ifnum\mgnf=0\special{#4.ps 1. 1. scale} \fi
\ifnum\mgnf=1\special{#4.ps 1.2 1.2 scale}\fi
\fi }\hfill \raise\yshift\hbox{#5}}}

\def\insertplotttt#1#2#3#4{\par%
\xwidth=#1 \xshift=\hsize \advance\xshift by-\xwidth \divide\xshift by 2%
\yshift=#2 \divide\yshift by 2%
\line{\hskip\xshift \vbox to #2{\vfil%
\ifnum\driver=0 #3
\special{ps: plotfile #4.ps} 
\fi
\ifnum\driver=1 #3 \includegraphics{#4.ps}\fi
\ifnum\driver=2 #3
\ifnum\mgnf=0\special{#4.ps 1. 1. scale} \fi
\ifnum\mgnf=1\special{#4.ps 1.2 1.2 scale}\fi
\fi }\hfill}}

\newdimen\xshift \newdimen\xwidth \newdimen\yshift
\def\eqfig#1#2#3#4#5{
\par\xwidth=#1 \xshift=\hsize \advance\xshift
by-\xwidth \divide\xshift by 2
\yshift=#2 \divide\yshift by 2
\line{\hglue\xshift \vbox to #2{\vfil
\ifnum\driver=0 #3
\special{ps: plotfile #4.ps} 
\fi
\ifnum\driver=1 #3 \includegraphics{#4.ps}\fi
\ifnum\driver=2 #3 \special{
\ifnum\mgnf=0 #4.ps 1. 1. scale \fi
\ifnum\mgnf=1 #4.ps 1.2 1.2 scale\fi}
\fi}\hfill\raise\yshift\hbox{#5}}}

\def\figini#1{
\def\8{\write13}
\catcode`\%=12\catcode`\{=12\catcode`\}=12
\catcode`\<=1\catcode`\>=2
\openout13=#1.ps}

\def\figfin{
\closeout13
\catcode`\%=14\catcode`\{=1
\catcode`\}=2\catcode`\<=12\catcode`\>=12
}

\let\0=\noindent

\def\ie{\hbox{\it i.e.\ }}

\ifnum\mgnf=0
\def\*{\vglue0.3truecm}
\fi
\ifnum\mgnf=1
\def\*{\vglue0.5truecm}
\fi

\let\0=\noindent
\def\\{\hfill\break}
\let\==\equiv

\let\io=\infty 

\def\tende#1{\,\vtop{\ialign{##\crcr\rightarrowfill\crcr
              \noalign{\kern-1pt\nointerlineskip}
              \hskip3.pt${\scriptstyle #1}$\hskip3.pt\crcr}}\,}
\def\otto{\,{\kern-1.truept\leftarrow\kern-5.truept\to\kern-1.truept}\,}

\def\VV{{\cal V}}

\def\TT{{\cal T}}\def\NN{{\cal N}}
\def\RR{{\cal R}}\def\LL{{\cal L}}

\def\T#1{{#1_{\kern-3pt\lower7pt\hbox{$\widetilde{}$}}\kern3pt}}
\def\VVV#1{{\underline #1}_{\kern-3pt
\lower7pt\hbox{$\widetilde{}$}}\kern3pt\,}
\def\W#1{#1_{\kern-3pt\lower7.5pt\hbox{$\widetilde{}$}}\kern2pt\,}
\def\Re{{\rm Re}\,}\def\Im{{\rm Im}\,}

\def\sign{{\rm sign}\,}
\def\indica{\leaders \hbox to 0.5cm{\hss.\hss}\hfill}
\def\guida{\leaders\hbox to 1em{\hss.\hss}\hfill}
\mathchardef\oo= "0521

\def\pp{{\bf p}}\def\xx{{\bf x}}
\def\yy{{\bf y}}\def\kk{{\bf k}}
\def\ie{{\it i.e. }}   
\newtoks\footline \footline={\hss\tenrm\folio\hss}
\let\ciao=\bye
\def\qed{\raise1pt\hbox{\vrule height5pt width5pt depth0pt}}

\ifnum\mgnf=0
\def\openone{\leavevmode\hbox{\ninerm 1\kern-3.3pt\tenrm1}}%
\fi
\ifnum\mgnf=1
\def\openone{\leavevmode\hbox{\ninerm 1\kern-3.63pt\tenrm1}}%
\fi
\def\indic{\hbox{\raise-2pt \hbox{\indbf 1}}}


\let\figini=\initfig
\let\figfin=\endfig

\figini{fig1}
\write13<
\write13<
\write13</origine1assexper2pilacon|P_2-P_1| { 
\write13<
\write13<4 2 roll 2 copy translate exch 4 1 roll sub 3 1 roll >
\write13<exch sub 
\write13<2 copy atan rotate 2 copy exch 4 1 roll mul >
\write13<3 1 roll mul add sqrt } def>
\write13<>
\write13</onda0 {
\write13<dup dup dup 10 div round div dup 100 div>
\write13<exch 1 exch div 1 mul 360 mul>
\write13<3 1 roll exch 0 3 1 roll 0 0 moveto >
\write13<{dup 2 index mul sin 4 mul lineto} >
\write13<for stroke pop pop } def >
\write13<>
\write13</punta0 { 
\write13<0 0 moveto dup dup 0 exch 2 div lineto 0 >
\write13<lineto 0 exch 2 div neg lineto>
\write13<0 0 lineto fill stroke } def>
\write13<>
\write13</onda { 
\write13<gsave origine1assexper2pilacon|P_2-P_1| >
\write13<onda0 grestore} def>
\write13<>
\write13</freccia0 { 
\write13<0 0 moveto 0 2 copy lineto stroke >
\write13<exch 2 div exch translate 7 punta0 >
\write13<stroke } def>
\write13<>
\write13</freccia{ 
\write13<gsave origine1assexper2pilacon|P_2-P_1| >
\write13<freccia0 grestore} def>
\write13<>
\write13</punto { gsave 
\write13<3 0 360 newpath arc fill stroke grestore} def>
\write13</tondo { gsave 
\write13<3 0 360 newpath arc stroke grestore} def>
\write13<>
\write13<>
\write13<gsave>
\write13<0 0 40 0 freccia>
\write13<40 0 punto>
\write13<40 0 40 40 onda>
\write13<40 0 80 0 freccia>
\write13<80 0 punto>
\write13<80 0 80 40 onda>
\write13<80 0 120 0 freccia>
\write13<120 0 punto>
\write13<120 0 120 40 onda>
\write13<120 0 160 0 freccia>
\write13<160 0 160 40 onda>
\write13<160 0 punto>
\write13<160 0 200 0 freccia>
\write13<grestore>
\figfin


\let\figini=\initfig
\let\figfin=\endfig

\figini{fig2}
\write13<
\write13<
\write13</origine1assexper2pilacon|P_2-P_1| { 
\write13<
\write13<4 2 roll 2 copy translate exch 4 1 roll sub 3 1 roll >
\write13<exch sub 
\write13<2 copy atan rotate 2 copy exch 4 1 roll mul >
\write13<3 1 roll mul add sqrt } def>
\write13<>
\write13</onda0 {
\write13<dup dup dup 10 div round div dup 100 div>
\write13<exch 1 exch div 1 mul 360 mul>
\write13<3 1 roll exch 0 3 1 roll 0 0 moveto >
\write13<{dup 2 index mul sin 4 mul lineto} >
\write13<for stroke pop pop } def >
\write13<>
\write13</punta0 { 
\write13<0 0 moveto dup dup 0 exch 2 div lineto 0 >
\write13<lineto 0 exch 2 div neg lineto>
\write13<0 0 lineto fill stroke } def>
\write13<>
\write13</onda { 
\write13<gsave origine1assexper2pilacon|P_2-P_1| >
\write13<onda0 grestore} def>
\write13<>
\write13</freccia0 { 
\write13<0 0 moveto 0 2 copy lineto stroke >
\write13<exch 2 div exch translate 7 punta0 >
\write13<stroke } def>
\write13<>
\write13</freccia{ 
\write13<gsave origine1assexper2pilacon|P_2-P_1| >
\write13<freccia0 grestore} def>
\write13<>
\write13</punto { gsave 
\write13<3 0 360 newpath arc fill stroke grestore} def>
\write13</tondo { gsave 
\write13<3 0 360 newpath arc stroke grestore} def>
\write13<>
\write13<>
\write13<gsave>
\write13<0 0 50 0 freccia>
\write13<50 0 punto>
\write13<50 0 50 50 onda>
\write13<50 0 100 0 freccia>
\write13<grestore>
\figfin

\openin14=\jobname.aux \ifeof14 \relax \else
\input \jobname.aux \closein14 \fi
\openout15=\jobname.aux
\null
\def\E{{\cal E}}
\centerline{\titolone Small denominators and
anomalous behaviour
in the}
\centerline{\titolone incommensurate Hubbard-Holstein model}
\vskip1.truecm
\centerline{{\titolo Vieri Mastropietro}\footnote{${}^\ast$}{\ottorm
Supported by MURST, Italy.}}
\centerline{Dipartimento di Matematica, Universit\`a di Roma ``Tor Vergata''}
\centerline{Via della Ricerca Scientifica, I-00133, Roma}
\vskip.2truecm
\line{\vtop{
\line{\hskip1.5truecm\vbox{\advance \hsize by -3.1 truecm
\0{\cs Abstract.}
{\it We consider a system of interacting fermions
on a chain in a periodic potential incommensurate
with the chain spacing.
We derive a convergent perturbative expansion,
afflicted by a small denominator problem and
based
on renormalization group, for
the two point 

Schwinger function.
We obtain
the large distance behavior of the
Schwinger function, which is anomalous and described by critical indices,
related to the gap and the wave function renormalization.
}} \hfill} }}
\vskip1.2truecm
\centerline{\titolo 1. Introduction.}
\*\numsec=1\numfor=1
\0{\bf 1.1} The {\it Holstein-Hubbard model}
describes a system of $d=1$ interacting spinless fermions
on a chain with unit spacing, moving in a periodic force
{\it incommensurate} with the spacing of the chain. The hamiltonian is
$$ H=H_0+u P+\l V \qquad H_0=\sum_{x,y\in\L}
t_{xy}\,\psi^+_{x}\psi^-_{y} \Eq(1.1)$$
$$P=\sum_{x\in\L} \f_x\psi^+_{x}\psi^-_{x}\quad
V=\sum_{x,y\in\L} v(x-y)\psi^+_{x}\psi^-_{x}\psi^-_{y}\psi^+_{y}$$

where $x,y$ are points on the one-dimensional lattice $\L$
with unit spacing, length $L$ and periodic boundary conditions; we shall
identify $\L$ with $\{x\in Z:\ -[L/2]\le x \le [(L-1)/2]\}$.
Moreover the matrix $t_{xy}$ is
defined as $t_{xy}=\d_{x,y}-(1/2)[\d_{x,y+1}+\d_{x,y-1}]$,
where $\d_{x,y}$ is the Kronecker delta.
The fields $\psi_x^{\pm}$ are creation ($+$) and annihilation ($-$)
fermionic fields, satisfying periodic boundary conditions:
$\psi_x^{\pm}=\psi_{x+L}^{\pm}$. We set $\xx=(x,t)$,
$-\b/2\le t \le \b/2$ for some $\b>0$; on $t$ antiperiodic boundary
conditions are imposed.

The term $P$ represents the interaction of the fermions with
a classical field. We are interested in studying potentials
which, in the limit $L\to\io$,
have the form $\f_x=\bar\f(2px)$, where $\bar\f$ is a
real function on the
real line $2\pi$-periodic and $p/\p$ is an irrational number, so
that the field has a period which is {\it incommensurate}
with the period of the lattice.
We also impose that $\bar\f(u)$ is of mean zero (its mean value can be
absorbed in the chemical potential), even and analytic in $u$,
so that
$$\bar\f(u) = \sum_{0\neq n \in Z}
\hat \f_n\,e^{inu} \; , \qquad |\hat \f_n| \le F_0 \, e^{-\x|n|}\; , \qquad
\hat \f_n=\hat \f_{-n}=\hat \f_n^* \; .
\Eq(1.2) $$
At finite volume we need a potential satisfying periodic boundary conditions;
hence, at finite $L$, we approximate $\f_x$ by
$$ \f^{(L)}_x = \sum_{n=-[L/2]}^{[(L-1)/2]}
\hat \f_n\,e^{2inp_Lx}\; , \Eq(1.13) $$
where $p_L$ tends to $p$ as $L\to\io$ and is of the form
$p_L=n_L\pi/L$, with $n_L$ an integer, relatively prime with respect to $L$.
The definition of $p_L$ implies that $2np_L$ is an allowed momentum
(modulo $2\p$). For technical reasons we need $p_L$
verifying the {\it Diophantine property} \equ(1.19) uniformly in $L$ and
in [BGM1], App. 1, a sequence of numbers verifying \equ(1.19)
is constructed. The thermodynamic limit
is then taken along a particular diverging sequence of volumes
and only along this sequence we can solve the problem (a non infrequent
situation in solid state physics, in which some models are solved only
with
particular boundary conditions).

The term $V$ in the hamiltonian represents the
interaction of the fermions by a short range two
body potential; in particular we assume $|v(x-y)|\le C e^{-\k |x-y|}$ for suitable
values of the constants $C,\k$.

Finally if $E_0^n$ is the {\it ground state energy} ,\ie the minimum value
of $H$ over the eigenstates $\psi_n$ with $n$ particles,
the {\it spectral gap}
is defined as $\D=E_0^{n+1}+E_0^{n-1}-2 E_0^n$. We denote moreover
the infinite volume zero
temperature
two point Schwinger function by
$\lim_{L,\b\to\io}S^{L,\b}(\xx;\yy)=S(\xx;\yy)$, defined in \equ(1.10f).
\vskip.5cm
\0{\bf 1.2} If $\l=0$ the Hamiltonian $H_0+u P\equiv \bar H$
is quadratic
in the Fermi fields and its
eigenfunctions
are the antisymmetrized product of the
one particles wavefunctions $\psi(x)$ of the
{\it finite difference Schroedinger  equation}
$$-\psi(x+1)-\psi(x-1)+u \f_x\psi(x)=E\psi(x)\Eq(1.3aa)$$
with $\f_x$ defined as above.
It is known that, for $u$ small enough
and if $p$ verifies a {\it diophantine condition}
$||2 n p||_{T^1}\ge C_0 |n|^{-\t}$,
for any $0\not= n\in Z$, there are, for particular values of $E$,
eigenfunctions which are {\it quasi Bloch waves}
of the form $\psi(x)=e^{-i k(E) x} U(k(E),x,u)$,
with $U(k(E),x,u)=\bar U(k(E),p x,u)$ and
$\bar U$ $2 \pi$-periodic in $px$.
In particular this is true if
$||k(E)+np||_{T^1}\ge C_0 |n|^{-\t}$,
$0\not =n\in Z$, [DS], or if
$k(E)=np$, $0\not= n\in Z$, [JM],[MP],[E] (strictly
speaking these results were proved for the almost
equivalent problem of the Schroedinger equation
in the continuum with a quasi
periodic potential, but
one can extend
them to this case, see [BLT]).
There are, for a generic potential,
{\it infinitely many
gaps} in the spectrum
in correspondence of the values
of $k(E)=np$ ${\rm mod.}\; 2\pi$, and
the spectrum is a Cantor set [E].
These results are obtained by KAM
iterative techniques, because the perturbative series
for the eigenvectors and eigenvalues of \equ(1.3aa)
are characterized by a
{\it small denominator problem} quite similar
to the one in the series for the invariant tori of
classical
hamiltonians close to integrable ones. It is useful to compare
the above results with their correspondent in the {\it commensurate
case}, in which $p\over\pi$ is a rational number.
In this case all the eigenfunctions
are Bloch waves, and the gaps are still in correspondence of
$k(E)=np, {\rm mod.}\; 2\pi$, but
the gaps are a finite number, for the rationality of $p/\pi$.
In the commensurate case the many body system
described by $\bar H$ has two physical phases.
For values of the chemical potential
$\mu$ (hence of the density) outside the gaps
of the one particle spectrum,
$S(\xx;\yy)$ decays with a power law for large
distances and the (infinite volume) ground state has no gaps $\D=0$;
this is the {\it metallic phase}.
Choosing $\mu$ in correspondence
of a gap in the one particle spectrum
the Schwinger function has a faster than any power
decay and the ground state
has a gap $\D\not=0$; this is the {\it insulating phase}.
In the incommensurate case the situation is more complex,
but still, see [BGM1],
for values of the chemical potential
corresponding to a gap
in the one particle spectrum ($\mu=1-\cos(mp_L)$, $m$ integer) and small
$u$,
$S(\xx;\yy)$ has a faster than any power
decay and the ground state has a gap $\D\not=0$. We can
call this a {\it quasi-insulating phase}.

If we consider also an interaction between fermions
we can expect that
still there is a {\it quasi-insulating} phase,
\ie for suitable values of the
chemical potential (hence of the density)
$S(\xx;\yy)$ has a faster than any power decay (but, as we
will see, {\it anomalous}) and
the ground state of $H$ has a gap $\D\not=0$.
However
there is in general no reason for which the chemical potentials
corresponding to the quasi-insulating phase
had to be the same as in the $\l=0$,
and indeed we find that they are different. The right
chemical potential to study the quasi-insulating phase
is unknown and it is part of the problem; it turns out that
$\mu=1-\cos(mp)-\nu$, with $\nu\equiv\nu(\l,u)$
is a suitable function such that
$\nu(0,u)=0$.
Defining the {\it Fermi momentum} as the momentum for which the
occupation number or some of its derivatives are singular,
our choice of the chemical potential corresponds to fix the {\it Fermi
momentum} of the interacting model as $p_F=mp$, $m\in Z^+$.
One of the main points of our analysis, as well as of the preceding
papers starting from [BG], is to use as a physical parameter the Fermi
momentum rather than the density. Of course if the formal
{\it Luttinger
theorem} (see for instance [BGL])
holds this is equivalent to fix the density;
but its validity is an open problem and there is no need to discuss it here.
Fixing the Fermi momentum is very reasonable in a work technically based
on renormalization group. Moreover this is the most natural choice from a
physical point of view, as discussed in the next section.
Nevertheless the problem of fixing the chemical potential to a $\l.u$
independent value is an interesting mathematical problem. To obtain
results for this problem from ours  has to solve an
implicit function problem (non trivial as our series are defined, as 
functions of $p_F$, only on $p_F$ verifying a diophantine condition),
but this will be not addresses here.
\vskip.5cm
\0{\bf 1.3}
Let we discuss an important physical application of our results,
so motivating our choice of fixing the Fermi momentum.
Peierls [P] and Fr\"ohlich [F] suggested that
one-dimensional metals are unstable at low temperature,
in the sense that they can lower
their energy through a periodic distorsion of the ``physical
lattice'' with period ${\pi\over p_F}$, where $p_F$ is the Fermi
momentum. Such a distorsion is called
{\it Charge Density Wave} (CDW),
as both the "physical lattice" and the
electrons
charge density form a new periodic structure with period
possibly bigger than the original lattice period $1$.
The CDW is usually
represented as a function $\bar\phi(2 p_F x)$.
Quite interesting is the case of {irrational} $p_F\over\pi$,
as Fr\"ohlich suggested that, if the period of the CDW
is {\it incommensurate} with the original period of the lattice,
the CDW has an arbitrary phase and so
it should carry an electric current.
The properties of many
compounds are explained in terms of
incommensurate CDW, see for instance
[L].
In recent times systems with an incommensurate CDW have been
reconsidered in the context of high-$T_c$
superconductivity; in particular it was suggested that it is crucial
to take into account also the interaction between fermions, see [A].
An interacting Fermi system with an incommensurate
CDW is then described by the Holstein-Hubbard model \equ(1.1) with
$p_F=p$, so our
results describe this physical situation.
We will see that the interaction introduces an anomalous behavior in the
model.

There is no mathematical proof that an incommensurate CDW really
exists also in presence of an interaction between fermions.
One has to
show that the ground state energy of \equ(1.1) as a function of
$\phi_x$ is minimized by $\phi_x=\bar\phi(2 p_F x)$. Up to now this was
proved only
if $\l=0$ and $p_F=\pi/2$; recent results [BGM2] show
that $\bar\phi(2 p_F x)$ is a stationary point for
the ground state energy of \equ(1.1), if $\l=0$ and $p_F/\pi$ rational.
It is clear, in any event that our results could represent a
starting point
for
this problem when $\l\not=0$, as from them one can write the ground state
energy
as a well defined expansion (\ie our work has the analogous role of [BGM1]
with respect to [BGM2]).
\vskip.5cm
\0{\bf 1.4}
Denote by $\|\a-\b\|_{T^1}$
the distance on $T^1$ of $\a,\b\in {T}^1$,
and, for $\xx=(x,x_0), \yy=(y,y_0)\in R^2$, by $|\xx-\yy|$
the distance $|\xx-\yy|=\sqrt{(x-y)^2+v_0 (x_0-y_0)^2}$, $v_0=\sin p_F$,
$p_F=\cos^{-1}(1-\mu)$.
The definition of the two-point Schwinger function $S^{L,\b}(\xx;\yy)$
is standard and it will be recalled below,
see \equ(1.10f). $\mu$ will be the chemical potential
of the $\l=0$ theory and $\bar\mu$
of the $\l\not=0$ theory.
Moreover $O(x,y,z)=O(|x|)+O(|y|)+O(|z|)$.

With the above definitions we shall prove the following theorem.
\vskip.5cm
\0{\cs Theorem}
{\it Let be $S^{L,\b}(\xx;\yy)$ the two point Schwinger function defined
in \equ(1.10f) with chemical potential $\bar\mu$.
Let us consider a sequence $L_i$, $i\in Z^+$, such that
$$ \lim_{i\to\io} L_i=\io \; , \qquad
\lim_{i\to\io}p_{L_i}=p \;  $$
Let be $\mu=1-\cos(\bar m p_{L_i})$, if $\bar m$ is a positive integer
such that $\hat\f_{\bar m}\not=0$, and
$p_{L_i}$ satisfies the {\sl diophantine condition}
$$ \|2n p_{L_i} \|_{T^1} \ge C_0 |n|^{-\t} \; , \qquad
0\neq n \in Z \;\qquad |n|\le {L_i\over 2} , \Eq(1.19) $$
for some positive constants $C_0$ and $\t$ independent of $i$.
Then there exists an $\e_0>0$ (independent from $i,\b$)
and four
functions $\nu\equiv\nu(\l,u)$, $\h_i(\l,u)$, $i=1,2,3$,
continuous for
$|u|,|\l|\le \e_0$ and
$\nu(\l,u)=O(\l)$, $\h_3(\l,u)=\h_1(\l,u)(1+\h_2(\l,u))^{-1}$ and
$\h_1(\l,u)=\b_1\l^2+\l^2 O(\l,u,\hat u)$,
$\h_2(\l,u)=\b_2\l+|\l|
O(\l,u,\hat u)$, with $\b_1,\b_2$ positive generically non vanishing
constants, such that, if
$\bar\mu=\mu-\nu(\l,u)$ the following propositions are
true.

\vskip.2truecm
(i) There exists the limit
$\lim_{\b\to\io \atop i\to\io} S^{L_i,\b}(\xx;\yy) = S(\xx;\yy)$
\vskip.2cm
(ii) $S(\xx;\yy)$ is continuous as a function of $\l,u$.
Moreover, defining
$$\hat u=|u\hat\f_{\bar m}|^{1+\h_2}\qquad \hat
Z=|u\hat\f_{\bar m}|^{-\h_1}\Eq(z.3)$$
for all $N$ there is $C_N$ such that for all
$|{\bf x} - {\bf y}| \ge \hat u^{-1}$
("large distance faster than
any power decay") 
$$|S(\xx,\yy)|\le {1\over \hat Z} {\hat u}
{C_N \over 1 +( |\hat u| \,|\xx-\yy|)^N}\Eq(z.1)$$

If $1\le |\xx-\yy|\le \hat u^{-1}$ one has
("transient slow decay")
$$|S(\xx,\yy)|\le {K_1\over |\xx-\yy|^{1+\h_3}}\Eq(z.2)$$
with $K_1>0$ constant and
$$S(\xx,\yy)={1\over |\xx-\yy|^{\h_3}}(g(\xx-\yy)+O(\l,u,\hat u)
C_2(\xx;\yy))\Eq(lutt)$$
with $g(\xx-\yy)=\lim_{\l\to 0, u\to
0}S(\xx;\yy)$ and $|C_2(\xx;\yy)|\le {K_1\over |\xx-\yy|}$
for $|\xx-\yy|\le \hat u^{-1/2}$.
\vskip.2truecm
(iii)For any $i$ there is a spectral gap $\D $ verifying
$$\D\geq {\hat u\over 2}\Eq(z.4)$$
}
\*
\0{\bf 1.5} The above theorem shows that also in presence of an
interaction between fermions there is a quasi-insulating phase; this is quite
remarkable as our results {\it hold also if $|u\hat \f_{\bar m}|<<|\l|$}
\ie if
the interaction between fermions is much larger that the amplitude of the
incommensurate potential.
Like in the non interacting $\l=0$
case, one can distinguish
two regions in the large distance behavior of the Schwinger function:
a transient slow decay and a long distance
faster than any power decay. In the first region there is still a
power law decay, but with a non universal exponent
$1+\h_3$ instead of $1$ ({\it anomalous behaviour});
in the second region the decay is still faster than
any power, like in the $\l=0$ case, but the decay rate
is $O(\hat u)$ instead of $O(u)$.
We think that the optimal bound in the large distance
behaviour is $|S(\xx-\yy)|\le {
c_1 e^{- c_2 \hat u |\xx-\yy|}\over
|\xx-\yy|^{1+\h_3} }$,
for some constants $c_i>0$,
and such bound could be possibly proved by a\ slight
improvement of our techniques. The variation of the decay rate
suggests that
the interacting ground state gap is $O(\hat u)$, \ie the ground
state has an
{\it anomalous gap}, and this is confirmed by
\equ(z.4), which
if the interaction is
attractive ($\l\le 0$)
says that the ratio between the bare
and interacting gap is
$<<1$ and diverging as $u\to 0$. The phase of the system
described by the above theorem
can be called {\it anomalous quasi-insulating phase}.

The above theorem shows that
the system with $\l\not=0$
is not "analitycally close" to the $\l=0$ one; the Schwinger
functions depends by $\l,u$ {\it not analytically} and this makes
necessary the use of renormalization group methods.
The Schwinger functions of the Holstein-Hubbard model
are studied writing them as a {\it functional
Grassmanian integral} which is
expressed by a perturbative series.
The small denominator problem afflicting
this
series is controlled, like in [BGM1], by using
a sort of {\it Bryuno Lemma} [B]
(see sec. 3.6 below) and using suitable {\it cancellations}
to face the problem of the {\it resonances}. Such cancellations are
implemented by renormalization group techniques.
The approach we follow
is then very closely related to the "direct"
methods developed in recent years
for proving the KAM theorem by showing the convergence of the {\it
Lindstedt series}
expressing the invariant tori, see [E1],[G1],[CF],[GM].
However in the above quoted
papers (including [BGM1]) the perturbative expansion can be expressed
in terms of {\it Feynmann graphs} which are only {\it tree graphs}
\ie with no loops:
in fact the KAM theorem is a classical problem
and in [BGM1] is discussed the hamiltonian \equ(1.1) with $\l=0$ which
defines a {\it non interacting model} \ie bilinear in the fields.
On the contrary the perturbative series for the Schwinger functions of
the Holstein-Hubbard model are expressed
in terms of Feynmann graphs with loops.
So a small denominator problem
in a fully interacting quantum theory is solved, and our work can be
considered a {\it quantum KAM theorem}. The main
idea is to combine such direct methods to study KAM problems with the
renormalization group techniques developed to study interacting fermions
starting from [BG] (similar techniques were introduced also in [FMRT]).

\vskip.5cm

\0{\bf 1.6} In order to make clearer
our results we state an immediate
consequence of the theorem proof, see sec. (4.5).
We can write
$$S(\xx;\yy)=S_1(\xx;\yy)+O(\l,u,\hat u) S_2(\xx;\yy)\; ,\Eq(1.21)$$
where, if $\kk=(k,k_0)$:
$$ S_1(\xx;\yy) = g^{(1)}(\xx;\yy) + \int {d\kk\over(2\p)^2}\,
[1-\hat f_1(\kk)]\, \phi(k,x,\hat u(k))\,\phi^*(k,y,\hat u(k)) \,
{e^{-ik_0(x_0-y_0)} \over -ik_0+\e(k,\hat u(k))}$$
with, if $p_F=\bar m p$
$$ \eqalignno{
g^{(1)}(\xx;\yy) & = \int {d\kk\over(2\p)^2}\,
{\hat f_1(\kk)\over \hat Z(k)}\, {e^{-ik_0(x_0-y_0)} \over -ik_0+ 1-\cos
k-\mu}\; ,\cr \e(k,\sigma) & = [1-\cos(|k|-p_F)] \cos p_F \cr
& +\sign (|k|-p_F) \, \sqrt{ [\sin(|k|-p_F)v_0]^2+\hat u(k)^2} \; , \cr
\phi(k,x,\hat u(k)) & =e^{-ikx} U(k,x,\hat u(k)) \; , & \eq(1.22) \cr
U(k,x,\hat u(k)) & = e^{i\sign(k)p_Fx}
\left[ \cos(p_Fx) \sqrt{ 1-
{\sign(|k|-p_F)\hat u(k) \over \sqrt{(\sin(|k|-p_F)v_0)^2+\hat u(k)^2}}}
\right. \cr
& \left. - i\,\sign(k)\sin(p_Fx) \sqrt{ 1+
{\sign(|k|-p_F)\hat u(k)\over\sqrt{(\sin(|k|-p_F)v_0)^2+\hat u(k)^2}}}
\; \right] \; . \cr} $$
Here $\hat f_1(\kk)$ denotes a cutoff function excluding
the two points $\kk=(\pm p_F,0)$, (defined in \equ(2.4))
and $\hat u(k),\hat Z(k)$ are two bounded
functions such that $|\hat u(k)-u|=O(u\l)|$,
$|\hat Z(k)^{-1}-1|=O(\l)$
for $||k|-p_F|>{p_F\over 2}$,
and  $\hat u(p_F)=\hat u, \hat Z(p_F)=\hat Z$ given by \equ(z.3); moreover
if $\l=0$ then $\hat u(k)=u$, $Z(k)=1$ (such functions will be explicitely
constructed in sec.(4.5)).
Finally $S_1(\xx,\yy), S_2(\xx,\yy)$ obeys to the same
bound (1.7),\equ(z.2).
In the $\l=0$ case $\phi(k,x,u)$ is the first order term
of a perturbative expansion for the quasi Bloch waves solving
\equ(1.3aa) with $k=p_F=\bar mp$, see [E], and,
as we can expect that $S_1$ is the dominant part of $S(\xx;\yy)$ for large
distances, this is in agreement with the results about
Schroedinger equation, see sec. (1.2).
If $\l\not=0$ again we can write $S(\xx;\yy)$ as sum of two terms,
and $S_1(\xx;\yy)$
is quite similar to the corresponding expression in the $\l=0$ case but
$\phi(k,x,u)$ is replaced by $\phi(k,x,\hat u(k))$
and there is a factor ${1\over \hat Z(k)}$ more.
It is natural to heuristically interpretate
this fact saying that the elementary exitation for small energy
in presence of interaction
are {\it quasi-particles} which are not
quasi-Bloch waves but
{\it
interacting quasi Bloch waves}
${1\over \sqrt{\hat Z(k)}}\phi(k,x,\hat u(k))$.

\vskip.5cm

\0{\bf 1.7} The Holstein-Hubbard model in the commensurate case
is discussed (essentially) in [BM2],[BM3], and one can see
that the Schwinger function has a similar behaviour.
Finally some comments
about the spin; the discussion is essentially identical to the one in
[BM2], [BM3] for the commensurate case and we do not repeat it.
In the spinning case the number
of running coupling constants is larger than in the spinless case and the
renormalization group flow is more complex. For repulsive interaction ($\l>0$)
things do not change while in the opposite case
the flow is unbounded and no conclusions
can be drawn (except if $|\hat\f_{\bar m}u|>e^{-1\over k|\l|}$, for a suitable
positive constant $k$ ); so our results are valid for spinning fermions only
if $\l>0$ or if $\l<0$ but $|\hat\f_{\bar m}u|>e^{-1\over k|\l|}$.

\vskip2.truecm
\centerline{\titolo 2. Multiscale decomposition and anomalous integration}
\*\numsec=2\numfor=1

\0{\bf 2.1} As it is well known, the Schwinger functions can be written as
power series in $\l$, convergent for $|\l| \le \e_\b$, for some constant
$\e_\b$ (the only trivial bound of $\e_\b$ goes to zero, as
$\b\to\io$). This power expansion is constructed in the usual way in terms of
Feynman graphs, by using as {\sl free propagator} the function
$$\eqalign{
g^{L,\b}(\xx;\yy) &\= g^{L,\b}(\xx-\yy)=
{{\rm Tr} \left[e^{-\b (H_0-\mu N)} {\bf T} (\psi^-_\xx \psi^+_\yy)\right] \over
{\rm Tr} [e^{-\b (H_0-\mu N)}]} = \cr
&={1\over L} \sum_{k\in {\cal D}_L}
e^{-ik(x-y)} \left\{ {e^{-\t e(k)} \over 1+e^{-\b e(k)}}
\indic(\t>0) - {e^{-(\b+\t) e(k)} \over 1+e^{-\b e(k)}} \indic(\t\le 0)
\right\}\; , \cr}\Eq(1.3f)$$
where $N=\sum_{x\in\L}\psi^+_x\psi^+_x$,
$\t=x_0-y_0$, $\indic(E)$ denotes the indicator function ($\indic(E)=1$, if
$E$ is true, $\indic(E)=0$ otherwise), $e(k)=1-\cos k -\mu$ and
${\cal D}_L\=\{k={2\pi n/L}, n\in Z, -[L/2]\le n \le [(L-1)/2]\}$.

It is easy to prove that, if $x_0\not= y_0$,
$$g^{L,\b}(\xx-\yy)= \lim_{M\to\io} {1\over L\b} \sum_{\kk\in {\cal D}_{L,\b}}
{e^{-i\kk\cdot(\xx-\yy)}\over -ik_0+\cos p_F-\cos k}\; , \Eq(1.4f)$$
where $\kk=(k,k_0)$, $\kk\cdot\xx=k_0x_0+kx$, ${\cal D}_{L,\b}\={\cal D}_L
\times {\cal D}_\b$, ${\cal D}_\b\=\{k_0=2(n+1/2)\pi/\b, n\in
Z, -M\le n \le M-1\}$ and $p_F$ is the {\sl Fermi momentum},
defined so that $\cos p_F =1-\mu$ and $0\le p_F \le \p$.

Hence, if we introduce a finite set of Grassmanian
variables $\{\psi^\pm_\kk\}$, one for each of the allowed $\kk$ values, and a
linear functional $P(d\psi)$ on the generated Grassmanian algebra, such that
$$\int P(d\psi) \psi^-_{\kk_1}\psi^+_{\kk_2} = L\b \d_{\kk_1,\kk_2}
\hat g_{\kk_1}\;,\quad \hat g_\kk= {1\over -ik_0+\cos p_F-\cos k}
\; ,\Eq(1.5f)$$
we have
$${1\over L\b} \sum_{\kk\in {\cal D}_{L,\b}} \, e^{-i\kk\cdot(\xx-\yy)} \,
\hat g_\kk = \int P(d\psi)\psi^-_\xx \psi^+_\yy
\= g^{L,\b}(\xx;\yy) \; ,\Eq(1.6f)$$
where the {\sl Grassmanian field} $\psi_\xx$ is defined by
$$\psi_\xx^{\pm}= {1\over L\b} \sum_{\kk\in {\cal D}_{L,\b}} \psi_\kk^{\pm}
e^{\pm i\kk\cdot\xx}\; .\Eq(1.7f)$$

The ``Gaussian measure'' $P(d\psi)$ has a simple representation in terms of
the ``Lebesgue Grassmanian measure'' $d\psi d\psi^+$,
defined as the linear
functional on the Grassmanian algebra, such that, given a monomial
$Q(\psi^-,\psi^+)$ in the variables $\psi_\kk^-,\psi_\kk^+$,
$$\int d\psi^- d\psi^+ Q(\psi^-,\psi^+) =
\cases{1& if $Q(\psi^-,\psi^+)=
\prod_\kk \psi^-_\kk\psi^+_\kk \; ,
$\cr 0& otherwise $\; .$\cr} \Eq(1.8f)$$
We have
$$P(d\psi) = \Big\{ \prod_\kk (L\b\hat g_\kk) \Big\}
\exp \Big\{-\sum_\kk (L\b \hat g_\kk)^{-1} \psi^+_\kk
\psi_\kk^- \Big\}
d\psi^- d\psi^+ \; .\Eq(1.9f)$$
Note that, since $(\psi_\kk^-)^2=(\psi^+_\kk)^2=0$,
$e^{-z \psi^+_\kk \psi_\kk}
=1-z \psi^+_\kk \psi_\kk$, for any complex $z$.

By using standard arguments (see, for example, [NO], where a different
regularization of the propagator is used), one can show that the Schwinger
functions can be calculated as expectations of suitable functions of the
Grassmanian field with respect to the ``Gaussian measure'' $P(d\psi)$.
In particular, the two-point Schwinger function
can be written, if $x_0\not= y_0$, as
$$ S^{L,\b}(\xx;\yy) = \lim_{M\to\io} {\int P(d\psi)\,
e^{-\VV(\psi)}\,\psi^-_{\xx}\psi^+_{\yy}
\over\int P(d\psi)\,e^{-\VV (\psi)}} \; , \Eq(1.10f) $$
where $\VV(\psi)=u P(\psi)+\l V(\psi)+\nu N(\psi)$ with
$$V(\psi)=\sum_{x,y\in\L}\int_{-\b/2}^{\b/2} dx_0\int_{-\b/2}^{\b/2} dy_0
v(x-y)\d(x_0-y_0)\psi_\xx^+\psi_\xx^-\psi_\yy^-\psi_\yy^+$$
$$ P(\psi)=\sum_{x\in\L} \int_{-\b/2}^{\b/2} dx_0
\Big[\f_x^{(L)}\psi_\xx^+ \psi^-_\xx \Big]\qquad
N(\psi)=\sum_{x\in\L} \int_{-\b/2}^{\b/2} dx_0
\psi_\xx^+ \psi^-_\xx      \; .\Eq(1.11f) $$

%
If $x_0=y_0$, $S^{L,\b}(\xx;\yy)$ must be defined as the limit of \equ(1.10f)
as $x_0-y_0\to 0^-$, as we shall understand always in the following.
\vskip.5cm
\0{\bf 2.2} We
start by evalutating the partition function\ie the denominator of \equ(1.10f)

$$\int P(d\psi) e^{-\VV(\psi)}\; , \Eq(2.1)$$

It is convenient to
decompose the Grassmanian integration $P(d\psi)$ into
a finite product of independent integrations:
$$ P(d\psi)=\prod_{h=h_\b }^1 P(d\psi^{(h)}) \; ,\Eq(2.1) $$

where $h_\b >-\io$ will be defined below (before \equ(2.9))
This can be done by setting
$$ \psi_\kk^{\pm}=\bigoplus_{h=h_\b }^1\psi_\kk^{(h)\pm} \; ,
\qquad \hat g_\kk=\sum_{h=h_\b }^1 \hat g^{(h)}_\kk  \; , \Eq(2.2) $$
where $\psi^{(h)\pm}_\kk$ are families of Grassmanian fields
with propagators $\hat g^{(h)}_\kk$ which are defined in the
following way.
%

We introduce a {\sl scaling parameter} $\g>1$ and a function
$\c(\kk') \in C^{\io}(T^1\times R)$, $\kk'=(k',k_0)$, such that,
if $|\kk'|\=\sqrt{k_0^2+||k'||_{T^1}^2}$:
$$ \c(\kk') = \c(-\kk') = \cases{
1 & if $|\kk'| <t_0 \= a_0/\g \;,$ \cr
0 & if $|\kk'| >a_0\; ,$\cr}\Eq(2.3)$$
where $a_0=\min \{p_F/2, (\p-p_F)/2 \}$. This definition
is such that the supports of $\c(k-p_F,k_0)$ and $\c(k+p_F,k_0)$ are
disjoint and the $C^\io$ function on $T^1\times R$
$$\hat f_1(\kk) \= 1- \c(k-p_F,k_0) - \c(k+p_F,k_0) \Eq(2.4)$$
is equal  to $0$, if $||k|-p_F||_{T^1}^2 +k_0^2<t_0^2$.

We define also, for any integer $h\le 0$,
$$f_h(\kk')= \c(\g^{-h}\kk')-\c(\g^{-h+1}\kk')\; ;\Eq(2.5)$$
we have, for any $\bar h<0$,
$$\c(\kk') = \sum_{h=\bar h+1}^0 f_h(\kk') +\c(\g^{-\bar h}\kk')\; .\Eq(2.6)$$
Note that, if $h\le 0$, $f_h(\kk') = 0$ for $|\kk'|
<t_0\g^{h-1}$ or $|\kk'| >t_0 \g^{h+1}$, and $f_h(\kk')=
1$, if $|\kk'| =t_0\g^h$.

We finally define, for any $h\le 0$:
$$ \hat f_h(\kk) = f_h(k-p_F,k_0) + f_h(k+p_F,k_0)\; ,\Eq(2.7) $$
$$ \hat g^{(h)}_\kk \= { \hat f_h(\kk) \over -ik_0+\cos p_F -\cos k}
\; . \Eq(2.8) $$

Note that, if $\kk\in {\cal D}_{L,\b}$, then $|k_0|\ge \p/\b$, implying that
$\hat f_h(\kk)=0$ for any $h< h_\b = \min \{h:t_0\g^{h+1} > \p/\b \}$.
Hence, if $\kk\in {\cal D}_{L,\b}$, the definitions \equ(2.4) and \equ(2.7),
together with the identity \equ(2.6), imply that
$$1=\sum_{h=h_\b }^1 \hat f_h(\kk) \; .\Eq(2.9)$$

The definition \equ(2.7) implies also that, if $h\le 0$, the support of
$\hat f_h(\kk)$ is the union of two disjoint sets, $A_h^+$ and $A_h^-$. In
$A_h^+$, $k$ is strictly positive and $||k-p_F||_{T^1}\le a_0\g^h \le a_0$,
while, in $A_h^-$, $k$ is strictly negative and
$||k+p_F||_{T^1}\le a_0\g^h$.
Therefore, if $h\le 0$, we can write $\psi^{(h)\pm}_{\kk}$ as the sum of two
independent Grassmanian variables $\psi_{\kk,\o}^{(h)\pm}$ with propagator
$$ \int P(d\psi^{(h)})\,\psi^{(h)-}_{\kk_1,\o_1}
\psi^{(h)+}_{\kk_2,\o_2} = L\b \d_{\kk_1,\kk_2}\,\d_{\o_1,\o_2}\,
\hat g^{(h)}_{\o_1}(\kk_1) \; , \Eq(2.10)$$
so that
$$ \psi^{(h)\pm}_{\kk}=\bigoplus_{\o=\pm 1}\psi^{(h)\pm}_{\kk,\o}
\; , \qquad \hat g^{(h)}_\kk=\sum_{\o=\pm 1} \hat g^{(h)}_\o(\kk) \; ,
\Eq(2.11) $$
$$ \hat g^{(h)}_\o(\kk)={\theta(\o k) \, \hat f_h(\kk) \over -ik_0
+ \cos p_F - \cos k }\; , \Eq(2.12) $$
where $\theta(k)$ is the (periodic) step function.
If $\o k> 0$, we will write in the following $k=k'+\o p_F$,
where $k'$ is the {\sl momentum measured from the Fermi surface} and we shall
define, if $h\le 0$,
$$ \tilde g^{(h)}_{\o}(\kk') \= \hat g^{(h)}_\o(\kk)=
{f_h(\kk') \over -i k_0+v_0 \o\sin k'+(1-\cos k')\cos p_F } \; , \Eq(2.13) $$
where $v_0=\sin p_F$. We call moreover
$\psi_{\kk,\omega}^{(\le k)\pm}=\bigoplus_{h=h_\b }^k\psi_{\kk,\omega}^{(h)\pm}$
and $g_{\kk,\omega}^{(\le k)}=\sum_{h=h_\b }^k \hat g^{(h)}_{\kk,\omega}$

This kind of decomposition is completely
standard in the theory of the $d=1$ Fermi system, starting from [BG];
we repeat it here only for clarity.

We define

$$e^{-\VV^{(0)}(\psi^{(\le 0)})}=\int P(d\psi^{(1)}) e^{-\VV^{(0)}(\psi^{(1)}+\psi^{(\le 0)})}$$

It is possible to prove , see [BGL], [BGPS] that

$$\VV^{(0)}(\psi^{(\le 0)})=\l {1\over (L\b)^4}\sum_{\kk_1,...,\kk_4\in
{\cal D}_{L,\b}} \hat v(\kk_1-\kk_2)
\psi^{(\le 0)+}_{\kk_1}
\psi^{(\le 0)-}_{\kk_2}\psi^{(\le 0)+}_{\kk_3}\psi^{(\le 0)-}_{\kk_4}
\d(\kk_1+\kk_3-\kk_2-\kk_4)$$
$$+ {1\over L\b}\sum_{\kk\in {\cal D}_{L,\b}}(\nu+F(\kk))
\psi^{(\le 0)+}_{\kk}\psi^{(\le 0)-}_{\kk}
+u\sum_{m=1}^\io \hat\phi_m {1\over L\b}\sum_{\kk\in {\cal D}_{L,\b}}
\psi^{(\le 0)+}_{\kk}
\psi^{(\le 0)-}_{\kk+2m \pp_L}+\psi^{(\le 0)+}_{\kk}
\psi^{(\le 0)-}_{\kk-2m \pp_L}$$
$$+\sum_{n=1}^\io\sum_{m=1}^\io {1\over (L\b)^n}\sum_{\kk_1,...,
\kk_n\in {\cal D}_{L,\b}}\psi^{(\le 0)\s_1}_{\kk_1}...
\psi^{(\le 0)\s_n}_{\kk_n}
W_{n,m}^{0}(\kk_1,...,\kk_n)\d(\sum_{i=1}^n\s_i\kk_i+2m\pp_L)$$
where $\sigma_i=\pm$, $|F(\kk)|\le C|\l|$ and the kernels $W_{n,m}^{0}(\kk_1,...,\kk_n;z)$ are $C^\io$ bounded functions such
that $W^0_{n,m}=W^0_{n,-m}$ and $|W^0_{n,m}|\le C^n z^{\max(2,n/2-1)}$ if
$z=Max(|\l|,u,|\nu|)$; moreover $\d(\kk)=L\b\d_{k_0}\d_{k}$ and
$\pp_L=(p_L,0)$.
\vskip.5cm
\0{\bf 2.3} The integration is performed iteratively, setting $Z_0=1$,
in the following way: once that the fields $\psi^0,...,\psi^{h+1}$ have
been integrated we have

$$\int P_{Z_h}(d\psi^{(\le h)}) \, e^{-\VV^{(h)}(\sqrt{Z_h}\psi^{(\le h)})}\Eq(3.22)$$

with, if $\pp_F=(p_F,0)$ and $C_h(\kk')^{-1}=\sum_{j=h_\b}^h f_j(\kk')$ and,
up to a constant:

$$ \eqalign{
 P_{Z_{h}}(d\psi^{(\le h)}) = &
\prod_{\kk}\prod_{\o=\pm1} d\psi^{(\le h)+}_{\kk'+\o\pp_F,\o}
d\psi^{(\le h)-}_{\kk'+\o\pp_F,\o} \cr
\exp \Big\{ &-\sum_{\o=\pm1} {1\over L\b} \sum_{\kk'\in
{\cal D}_{L,\b}} \,
C_h(\kk') Z_{h}
\Big[\Big( -ik_0-(\cos k'-1)\cos p_F +\o v_0\sin k' \Big) \cr
& \psi^{(\le0)+}_{\kk'+\o\pp_F,\o}
\psi^{(\le0)-}_{\kk'+\o\pp_F,\o}
+ \sigma_{h}(\kk') \, \psi^{(\le0)+}_{\kk'+\o\pp_F,\o}
\psi^{(\le0)-}_{\kk'-\o\pp_F,-\o} \Big]
\Big\} \; \cr} \Eq(3.10a) $$

It is convenient, for reasons which will be clear below, to split
$\VV^{(h)}$ as $\LL \VV^{(h)}+\RR \VV^{(h)}$, with $\RR=1-\LL$ and
$\LL$, the {\it localization operator}, is a linear operator defined
in the following way:

\vskip.5cm
1) If $n>4$ then
$$\LL\{{1\over (L\b)^n}\sum_{\kk'_1,...,\kk'_n\in {\cal D}_{L,\b}}
W_{n,m}^{h}(\kk'_1+\o_1
\pp_F,...)\prod_{i=1}^n\psi^{(\le h)\s_i}_{\kk'_i+\o_i \pp_F,\o_i}
\d(\sum_{i=1}^n \s_i(\kk'_i+\o_i \pp_F)+2m\pp_L)\}=0$$
\vskip.5cm
2) If $n=4$ then
$$\LL \{ {1\over (L\b)^4}\sum_{\kk'_1,...,\kk'_4\in {\cal D}_{L,\b}}
W_{4,m}^{h}(\kk'_1+\o_1 \pp_F,...)[\prod_{i=1}^4
\psi^{(\le h)\s_i}_{\kk'_i+\o_i \pp_F,\o_i}]
\d(\sum_{i=1}^4 \s_i(\kk'_i+\o_i \pp_F)+2m\pp_L)\}\Eq(loc1)$$
$$=\d_{\sum_{i=1}^4\sigma_i\o_i p_F+2m p_L,0}
{1\over (L\b)^4}\sum_{\kk'_1,...,\kk'_4\in {\cal D}_{L,\b}}
W_{4,m}^{h}(\o_1  \pp_F,...,
\o_4 \pp_F)
[\prod_{i=1}^4
\psi^{(\le h)\s_i}_{\kk'_i+\o_i \pp_F,\o_i}]\d(\sum_{i=1}^4
\s_i \kk'_i)$$
\vskip.5cm
3)If $n=2$ then
$$\LL\{ {1\over (L\b)^2}\sum_{\kk'_1,\kk'_2\in {\cal D}_{L,\b}}
W_{2,m}^{h}(\kk'_1+\o_1 \pp_F,
\kk'_2+\o_2 \pp_F)[\prod_{i=1}^2\psi^{(\le 0)\s_i}_
{\kk'_i+\o_i \pp_F,\o_i}]$$
$$\d(\sum_{i=1}^2 \s_i(\kk'_i+\o_i \pp_F)+2m\pp_L)\}
=\d_{(\o_1-\o_2)p_F+2m p_L,0}{1\over (L\b)}
\sum_{\kk'\in {\cal D}_{L,\b}}
 [W_{2,m}^{h}(\o_1 \pp_F,\o_2 \pp_F)+\Eq(loc2)$$
$$+\o_1E(k'+\o_1 p_F)\partial_{k}
W_{2,m}^{h}(\o_1 \pp_F,\o_2 \pp_F)+
k^0 \partial_{k_0}W_{2,m}^{h}(\o_1 \pp_F,\o_2 \pp_F)]
[\prod_{i=1}^2\psi^{(\le h)\s_i}_{\kk'_i+\o_i \pp_F,\o_i}]$$
\vskip.5cm
where $E(k'+\o p_F)=v_0\o\sin k'+(1-\cos k')\cos p_F$ and the symbol
$\partial_k,\partial_{k_0}$ means discrete derivatives. Note that
the r.h.s of \equ(loc1),\equ(loc2) are vanishing unless
$\sum_i \s_i\o_ip_F+2m p_L=0$ ${\rm mod.} 2\pi$.
\vskip.5cm
We can write then $\LL \VV^{(h)}$ in the
following way:

$$\LL\VV^{(h)}(\psi)=\g^h n_h F_\nu^{(\le h)}+s_h F_\s^{(\le h)}+z_h F_\z^{(\le h)}+a_h
F_\a^{(\le h)}+
i_h F_\iota^{(\le h)}+t_h F_\t^{(\le h)}+l_h F_\l^{(\le h)}
\;  \Eq(3.11a) $$
where

$$F_\nu^{(\le h)}=\sum_{\o=\pm 1} {1\over (L\b)}\sum_{\kk'\in {\cal D}_{L,\b}}
\psi^{(\le h)+}_{\kk'+\o \pp_F,\o}
\psi^{(\le h)-}_{\kk'+\o \pp_F,\o}$$

$$F_\s^{(\le h)}=\sum_{\o=\pm 1} {1\over (L\b)}\sum_{\kk'\in {\cal D}_{L,\b}}
\psi^{(\le h)+}_{\kk'+\o \pp_F,\o}
\psi^{(\le h)-}_{\kk'-\o \pp_F,-\o}$$

$$F_\a^{(\le h)}=\sum_{\o=\pm 1} {1\over (L\b)}\sum_{\kk'\in {\cal D}_{L,\b}}
E(k'+\o p_F) \psi^{(\le h)+}_{\kk'+\o \pp_F,\o}
\psi^{(\le h)-}_{\kk'+\o \pp_F,\o}$$

$$F_\z^{(\le h)}=\sum_{\o=\pm 1} {1\over (L\b)}\sum_{\kk'\in {\cal D}_{L,\b}}
(-i k_0) \psi^{(\le h)+}_{\kk'+\o \pp_F,\o}
\psi^{(\le h)-}_{\kk'+\o \pp_F,\o}$$

$$F_\iota^{(\le h)}=\sum_{\o=\pm 1}
{1\over (L\b)}\sum_{\kk'\in {\cal D}_{L,\b}} E(k'+\o p_F)
\psi^{(\le h)+}_{\kk'+\o \pp_F,\o}
\psi^{(\le h)-}_{\kk'-\o \pp_F,-\o}$$

$$F_\t^{(\le h)}=\sum_{\o=\pm 1} {1\over (L\b)}\sum_{\kk'\in {\cal D}_{L,\b}}  (-i k_0)
\psi^{(\le h)+}_{\kk'+\o \pp_F,\o}
\psi^{(\le h)-}_{\kk'-\o \pp_F,-\o}$$

$$F_\l^{(\le h)}={1\over (L\b)^4}\sum_{\kk'_1,...,\kk'_4\in {\cal D}_{L,\b}}
\psi^{(\le h)+}_{\kk'_1+\pp_F,1}
\psi^{(\le h)+}_{\kk'_1-\pp_F,-1} \psi^{(\le 0)-}_{\kk'_3+\pp_F,1}
\psi^{(\le 0)-}_{\kk'_4-\pp_F,-1}\d(\sum_{i=1}^4\s_i\kk_i)$$

and $\l_0=\l(\hat
v(0)-\hat v(2 p_F))+O(\l^2)$,
$s_0=u\hat\phi_{\bar m} +O(u\l^2)$, $t_0,i_0=O(u\l^2)$,
$a_0,z_0=O(\l^2)$,
$n_0=\nu+O(\l^2)$. This follows from the antisimmetry properties
of the Grassman variables and the Kroneker deltas in the r.h.s.
of \equ(loc1),\equ(loc2).

We write, if $\NN_h$ is a constant
$$\int P_{Z_h}(d\psi^{(\le h)}) \, e^{-\VV^{(h)}(\sqrt{Z_h}\psi^{(\le
h)})}={1\over\NN_h}
\int \tilde P_{Z_{h-1}}(d\psi^{(\le h)}) \, e^{-\tilde\VV^{(h)}(\sqrt{Z_h}\psi^{(\le h)})}
\Eq(3.22a)$$

where
$$ \eqalign{
&\tilde  P_{Z_{h-1}}(d\psi^{(\le h)}) =
\prod_{\kk}\prod_{\o=\pm1} d\psi^{(\le h)+}_{\kk'+\o\pp_F,\o}
d\psi^{(\le h)-}_{\kk'+\o\pp_F,\o} \cr
&\exp \Big\{ -\sum_{\o=\pm1} {1\over L\b} \sum_{\kk'\in {\cal D}_{L,\b}} \,
C_h(\kk') Z_{h-1}(\kk')
\Big[\Big( -ik_0-(\cos k'-1)\cos p_F +\o v_0\sin k' \Big) \cr
& \psi^{(\le0)+}_{\kk'+\o\pp_F,\o} \psi^{(\le0)-}_{\kk'+\o\pp_F,\o}
+ \sigma_{h-1}(\kk') \, \psi^{(\le0)+}_{\kk'+\o\pp_F,\o}
\psi^{(\le0)-}_{\kk'-\o\pp_F,-\o} \Big] \Big\} \; , \cr} \Eq(3.10b) $$

with $Z_{h-1}(\kk')=Z_h(1+C_h^{-1}(\kk')z_h)$, $Z_{h-1}(\kk')
\s_{h-1}(\kk')=Z_h(
\s_h(\kk')+C_h^{-1}(\kk') s_h)$ and $\tilde
\VV^{(h)}=\LL\tilde\VV^{(h)}+(1-\LL)\VV^{(h)}$ with

$$\LL\tilde\VV^{(h)}(\psi)=\g^h n_h F_\nu^{(\le h)}+(a_h-z_h)
F_\a^{(\le h)}+i_h F_\iota^{(\le h)}+t_h F_\t^{(\le h)}+l_h F_\l^{(\le h)}
\; . \Eq(3.11b)$$

The r.h.s of \equ(3.22a) can be written as
$$ {1 \over \NN_h}\int P_{Z_{h-1}}(d\psi^{(\le h-1)}) \int \tilde
P_{Z_{h-1}}(d\psi^{(h)}) \, e^{-\tilde \VV^{(h)}(\sqrt{Z_h}\psi^{(\le h)})} \; , \Eq(3.12a) $$
where $ P_{Z_{h-1}}(d\psi^{(\le h-1)})$ and $\tilde P_{Z_{h-1}}(d\psi^{(h)})$ are given
by \equ(3.10b) with $Z_{h-1}(\kk')$ replaced by $Z_{h-1}(0)\equiv Z_{h-1}$
and
$C_h(\kk')$ replaced with
$C_{h-1}(\kk')$
and $\tilde f_h^{-1}(\kk')$ respectively, if
$$\tilde f_h(\kk')=Z_{h-1}[{C_h^{-1}(\kk')\over Z_{h-1}(\kk')}-
{C_{h-1}^{-1}(\kk')\over Z_{h-1}}]$$
and $\psi^{(\le h)}$ replaced with
$\psi^{(\le h-1)}$ and $\psi^{(h)}$ respectively. Note that $\tilde f_h(\kk')$
is a compact support function, with support of width $O(\g^h)$
and far $O(\g^h)$ from the "singularity" \ie $\pp_F$.

The Grassmanian integration $\tilde P_{Z_{h-1}}(d\psi^{(h)})$
has propagator
$$ {g^{(h)}(\xx;\yy)\over Z_{h-1}} = \sum_{\o,\o'=\pm1}
e^{-i(\o x - \o' y)p_F}\,
{g^{(h)}_{\o,\o'}(\xx;\yy) \over Z_{h-1}}\; , \Eq(3.13a) $$
if
$$ {g^{(h)}_{\o,\o'}(\xx;\yy)\over Z_{h-1}}=\int \tilde
P_{Z_{h-1}}(d\psi^{(h)})\,
\psi^{(h)-}_{\xx,\o}\psi^{(h)+}_{\yy,\o'} \Eq(3.14) $$
is given by
$$ g^{(h)}_{\o,\o'}(\xx;\yy)={1\over L\b} \sum_{\kk'\in {\cal D}_{L,\b}} \,
e^{-i\kk'\cdot(\xx-\yy)} \tilde f_h(\kk')
[T_{h}^{-1}(\kk')]_{\o,\o'}
\; , \Eq(3.15a) $$
where the $2\times2$ matrix $T_{h}(\kk')$ has elements
$$ \cases{
[T_{h}(\kk')]_{1,1} =
\left(-ik_0-(\cos k'-1)\cos p_F+ v_0\sin k' \right) \; , & \cr
[T_{h}(\kk')]_{1,2} = [T_{h}(\kk')]_{2,1} =  \sigma_{h-1}(\kk') \; , & \cr
[T_{h}(\kk')]_{2,2} =
\left(-ik_0-(\cos k'-1)\cos p_F-v_0\sin k'\right) \; , \cr} \Eq(3.16a) $$
which is well defined on the support of $\tilde f_h(\kk')$, so that,
if we set
$$ A_{h}(\kk') = \det T_h(\kk') =
[ -ik_0-(\cos k'-1)\cos p_F ]^2 - (v_0\sin k')^2
- [\sigma_{h-1}(\kk')]^2 \; ,\Eq(3.17) $$
then
$$ T_{h}^{-1}(\kk')= {1\over A_{h}(\kk') }
\left( \matrix{
[\t_{h}(\kk')]_{1,1} & [\t_{h}(\kk')]_{1,2} \cr
[\t_{h}(\kk')]_{2,1} & [\t_{h}(\kk')]_{2,2} \cr} \right) \; , \Eq(3.18a) $$
with
$$ \cases{
[\t_{h}(\kk')]_{1,1} = \left[-ik_0-(\cos k'-1)
\cos p_F-v_0\sin k'\right] \; , & \cr
[\t_{h}(\kk')]_{1,2} = [\t_{h}(\kk')]_{2,1} =
-\sigma_{h-1}(\kk') \; , & \cr
[\t_{h}(\kk')]_{2,2} = \left[-ik_0-(\cos k'-1)\cos p_F
+ v_0 \sin k' \right] \; . & \cr} \Eq(3.19a) $$

Note that $\sigma_h(\kk')$ is a smooth function on $T^1\times R$ and, if $\kk'$
varies in the support of $C_h^{-1}(\kk')$ then there exists two positive constants
$c_1,c_2$ such that:
$$c_1 \sigma_h\le \sigma_h(\kk')\le c_2\sigma_h\Eq(ass)$$
if $\sigma_h\equiv \sigma_h(0)$, as $C_k^{-1}(\kk')=1$ for $k\ge h+1$.

The large distance behaviour of the propagator \equ(3.13a)
is given by the following lemma, see [BM2]:

\vskip.5cm
\0{\bf 2.4}{\rm LEMMA} {\it The propagator $g^{(h)}_{\o,\o'}(\xx;\yy)$
can be written as:
$$g^{(h)}_{\o,\o}(\xx;\yy)=g^{(h)}_{L;\o}(\xx;\yy)+
C_1^{(h)}(\xx;\yy)+C_2^{(h)}(\xx;\yy)
\Eq(bb)$$ with
$$g^{(h)}_{L;\o}(\xx;\yy)=\int d\kk {e^{-i\kk(\xx-\yy)}\over -i k_0+\o v_0 
k'}
\tilde f_h(\kk')$$
For any integer $N>1$ and for $|x-y|\le {L\over 2}$, $|x_0-y_0|\le
{\b\over 2}$ it holds
$|C_1^{(h)}(\xx;\yy)|\le {\g^{h}(\max \{\g^h,L^{-1}\}) C_N\over
1+(\g^h(\xx-\yy))^N}$
and $|C_2^{(h)}(\xx;\yy)|\le |{\sigma^h\over \g^h}|^2{(\max
\{\g^h,L^{-1}\}) C_N\over
1+(\g^h(\xx-\yy))^N}$.

Moreover
$$|g^{(h)}_{\o,-\o}(\xx;\yy)|\le |{\sigma^h\over \g^h}| {(\max
\{\g^h,L^{-1}\}) C_N\over
1+(\g^h(\xx-\yy))^N}$$} \vskip.5cm
Note that $g^{(h)}_{L;\o}(\xx;\yy)$ coincides with the propagator
"at scale $\g^h$" of the Luttinger model, see [BeGM]. This remark will be
crucial
for studying the Renormalization group flow, see sec.4. For the proof of
the bounds, one can proceed as in the proof of lemma 2.8 below.
\vskip.5cm
\0{\bf 2.5} We {\it rescale} the fields so that
$${1 \over \NN_h}\int P_{Z_{h-1}}(d\psi^{(\le h-1)}) \int \tilde
P_{Z_{h-1}}(d\psi^{(h)})
\, e^{-\hat\VV^{(h)}
(\sqrt{Z_{h-1}}\psi^{(\le h)})}$$
so that
$$\LL\hat\VV^{(h)}(\psi)=
\g^h\nu_h F_\nu^{(\le h)}+\d_h
F_\a^{(\le h)}+\iota_h F_\iota^{(\le h)}+\t_h F_\t^{(\le h)}+\l_h F_\l^{(\le h)}
\; . \Eq(3.11ah) $$
where by definition
$$\nu_h={Z_h\over Z_{h-1}}n_h;\quad
\d_h={Z_h\over Z_{h-1}}(a_h-z_h);\quad \t_h={Z_h\over Z_{h-1}} t_h;\quad \iota_h=
{Z_h\over
Z_{h-1}}i_0;\quad \l_h=({Z_h\over Z_{h-1}})^2 l_h$$

We call the set $\vec v_h=(\nu_h,\d_h,\t_h,\iota_h,\l_h)$ {\it running
coupling constants}.

We perform the integration
$$ \int \tilde P_{Z_{h-1}}(d\psi^{(h)}) \, e^{-\hat\VV^{(h)}
(\sqrt{Z_{h-1}}\psi^{(\le h)})}
= e^{-\VV^{h-1}(\sqrt{Z_{h-1}}\psi^{(\le h-1)}) + \tilde E_h} \; , \Eq(3.20h) $$
where $\tilde E_h$ is a suitable constant and
$$\LL \VV^{(h-1)}(\psi)= \g^{h-1} n_{h-1} F_\nu^{\le (h-1)}+s_{h-1}
F_\sigma^{\le (h-1)} +a_{h-1} F_\a^{(\le h-1)}+$$
$$z_{h-1} F_\z^{(\le h-1)}+i_{h-1} F_\iota^{(\le h-1)}+ t_{h-1} F_\t^{(\le
h-1)}+ l_{h-1} F_\l^{(\le -1)}\; , \Eq(3.21h) $$
\vskip.5cm
Note that the above procedure allows us to write the running coupling
constants $\vec v_h$ in terms of $\vec v_k$, $k\ge h+1$
$$\vec v_h=\vec\b(\vec v_{h+1},...,\vec v_{0})\Eq(beta)$$
The function $\vec\b(\vec v_{h+1},...,\vec v_{0})$ is
called {\it Beta function}.

The {\it effective potential} $\VV^{(h)}(\psi)$ is a sum of terms of the form
$${1\over (L\b)^n} \sum_{\kk'_1,...,\kk'_n
\in {\cal D}_{L,\b}} \d(\sum_{i=1}^n \sigma_i(\kk'_i+\o_i \pp_F)+2m\pp_L)
W_{n,m}^{h}(\kk'_1,..,\kk'_n;'_n;\{\o\})\prod_{i=1}^n \psi^{\sigma_i
(\le h)}_{\kk'_i+\o_i \pp_F,\o_i}\Eq(bbb)$$

\vskip.5cm
\0{\bf 2.6} Let we explain the main motivations of the integration
procedure discussed above. In a renormalization group
approach one has to identify the relevant, marginal and irrelevant
interactions. By a power counting argument
one sees
that the terms bilinear in the fields are relevant
and the quartic terms (or the bilinear
ones with a derivative respect to
some coordinate acting on the fields) are marginal.
However there are $O(L)$ many different terms
bilinear or quartic in the fields, depending on the value of $m$
in \equ(bbb), and so it seems that there are $O(L)$ different
running coupling constants, and their renormalization group flow
seems impossible to study. However it will turn out
that only a small subset of bilinear or quartic interactions are
really relevant or marginal: the terms such that
$\sum_i\s_i\oo_i p_F+2 m p_L=0$ ${\rm mod.}\; 2\pi$.
Then we define the $\LL$
operator \equ(loc1),\equ(loc2) just to extract such
terms from the effective potential. The reason why the terms
such that $\sum_i\s_i\oo_i p_F+2 m p_L\not=0$ are irrelevant
is quite clear in the commensurate case \ie if
$p/\pi$ is a rational number. In this case in fact
$||\sum_{i=1}^n \s_i\oo_i
p_F+2mp_L||_{T^1}$ is greater than some positive number, and this,
by the momentum conservation, {\it means that the momenta of the
fields cannot be
all at the same time closer
than some fixed quantity to the
singularity of the propagator}.
In the incommensurate case however due to
irrationality of $p/\pi$ the factor $||\sum_{i=1}^n \s_i\oo_i
p_F+2mp_L||_{T^1}$ can be as small as one likes for very large $m$
so that the momenta of the field
can be also very
close to the singularity. However by the diophantine condition
this will happen only for very large $m$, see Lemma 2.7 below,
and using the exponential decay of $\hat\f_m$ we will see that these terms are
indeed irrelevant.

The relevant terms are of two kinds;
the $\nu$ terms, reflecting
the renormalization of the Fermi momentum, and the $\s$ terms, related
to the presence of a gap in the spectrum. The first kind of terms are
faced in a standard way [BG] fixing properly the counterterm $\nu$
in the hamiltonian \ie fixing properly the chemical potential.
On the contrary there are
no
free parameters for the $\s$ terms in the hamiltonian and one has to proceed
in
a different way. One can naively think to perform a Bogolubov transformation,
so considering as free propagator a propagator corresponding to a
theory with a gap $O(u)$ at the Fermi surface.
This is essentially what one does in the $\l=0$ case, see [BGM1],
but here this does not work as the gap is deeply renormalized
by the interaction and {\it one has to perform
different Bogolubov transformations at each integration}.
In fact $\s_h$ is a sort of "mass terms"
with a non trivial renormalization group flow.

Regarding the
marginal terms, there is an anomalous wave function
renormalization which one has to take into account, what is expected as if
$u=0$ the theory is a Luttinger liquid.
In general the flow of the marginal terms can be controlled
using some cancellations due to the fact that the Beta function
is "close" (for small $u$) to the Luttinger model Beta function. In lemma 2.4
we write the propagator as the Luttinger model propagator
plus a remainder, so that the Beta function is equal to the Luttinger
model Beta function plus a "remainder" which is small if $\s_h\g^{-h}$
is small.

\vskip.5cm
\0{\bf 2.7} {\rm LEMMA} {\it Assume that
$\sum_{i=1}^n \sigma_i \o_i p_F+2mp_L\not=0$ ${\rm mod.}\; 2\pi$. Then
the contributions to $\VV^{(h)}$ of the form \equ(bbb) are vanishing
unless
$$|m|\geq {\cal A} [{\g^{-h\over\t}\over n^{1/\t}}-\bar m n]\Eq(3.21hh)$$
if ${\cal A}$ is a suitable positive constant}
\vskip.5cm
{\rm  PROOF} Remembering the compact support of the Grassmanian operator
$\psi^{\sigma_i (\le h)}_{\kk'_i+\o_i \pp_F,\o_i}$
we can write, using \equ(bbb)
$$a_0 n\g^h\geq ||\sum_{i=1}^n \sigma_ik'_i||_{T^1}
\geq ||2mp_L+\sum_{i=1}^n\sigma_i
\o_i p_F||_{T^1}\ge C_0 ( 2|n|\bar m+m)^{-\t}$$
from which \equ(3.21hh) follows.
\qed
\vskip.5cm
Let be
$$h^*={\rm inf}\{h\ge h_\b:a_0\g^{h+1}\ge 4|\sigma_{h}|\}\Eq(h*)$$
From Lemma 2.4 it follows trivially:
\vskip1cm

\0{\bf 2.8} {\rm LEMMA} {\it
For $h> h^*$ for any integer $N>1$ it is possible to find
a $C_N$ such that, for $|x-y|\le{L\over 2}$, $|x_0-y_0|\le{\b\over 2}$
$$|g^{(h)}_{\o,\o'}(\xx;\yy)|\le {C_N (\max \{\g^h,L^{-1}\})\over
1+(\g^h|\xx-\yy|)^N}$$}

{\rm Proof} For any $N\ge 0$, there is a constant $G_N$
such that, if $0\ge h\ge h^*$, $N_0,N_1\ge 0$ and $N_0+N_1=N$,
$$ \Big| D_0^{N_0} D_1^{N_1} \tilde g_{\o,\o'}^{(h)}(\kk') \Big|
\le G_N \g^{-hN}\,
{ \max\{\g^{h},|\s_{h}| \} \over \g^{2h} + \s_{h}^2 } \; , \Eqa(A2.1) $$
where $D_0$ and $D_1$ denote the discrete derivative with respect to $k_0$
and $k'$, respectively.

Hence, if $|x_0-y_0| \le \b/2$ and $|x-y|\le L_i/2$, we have
$$ \eqalign{
& \left( {\sqrt{2}\over \p} \right)^N |x_0-y_0|^{N_0} |x-y|^{N_1}
\Big| g^{(h)}(\xx;\yy) \Big| \le \cr
&\le \left| {\b\over 2\p} [e^{-i{2\p\over \b}(x_0-y_0)}-1] \right|^{N_0}
\left| {L_i\over 2\p} [e^{-i{2\p\over L_i}(x-y)}-1] \right|^{N_1}
\Big| g^{(h)}(\xx;\yy) \Big| = \cr
& = \Big| \sum_{\o,\o'=\pm1} e^{-i(\o x-\o' y)p_F}
{1\over L_i\b} \sum_{\kk\in {\cal D}_{L_i,\b}} \, e^{-i\kk'\cdot(\xx-\yy)}
\, D_0^{N_0} D_1^{N_1} \tilde g^{(h)}_{\o,\o'}(\kk') \Big| \;\le\cr
& \le C_N \g^h (\max \{\g^h,L^{-1}\}) \,\g^{-hN} {\max\{\g^h,|\s_h|\}
\over \g^{2h}+\s_h^2} \le C_N \g^{-hN}(\max \{\g^h,L^{-1}\})\; , \cr}
\Eqa(A2.2) $$
where $C_N$ denotes a varying constant, depending only on $N$, and
the factor $(\max \{\g^h,L^{-1}\})$ arises from the sum over $\kk'$ (note
that
the sum over $k_0$ always gives a factor $\g^h$, since $h_\b\le h^*\le
h$).
Therefore we have
$$ \Big| g^{(h)}(\xx;\yy) \Big| \le
{C_N \max \{\g^h,L^{-1}\} \over 1 + \g^{hN}|\xx-\yy|^N} \; .\Eqa(A2.3) $$

\vskip.5cm
\0{\bf 2.9} In the next section we will see that, using the above lemmas
and assuming
that the running coupling constants are bounded, the integrations (2.24)
are well defined for $0\ge h>h^*$.

The integration of the scale from $h^*$ to
$h_\b$ can be performed "in a single step"

$$\int P_{Z_{h^*}}(d\psi^{(\le h^*)})
e^{-\VV^{h^*}(\sqrt{Z_{h^*}}\psi^{(\le h^*)}}={1\over\NN_{h^*}}
\int \tilde P_{Z_{h^*-1}}(d\psi^{(\le h^*)})
e^{-\tilde\VV^{h^*}(\sqrt{Z_{h^*}}\psi^{(\le h^*)})}\Eq(uu)$$

Calling $\int \tilde P_{Z_{h^*-1}}(d\psi^{(\le h^*)})\psi_\xx^{+(\le
h^*)}\psi_\yy^{-(\le h^*)}\equiv
{g^{(\le h^*)}(\xx;\yy)\over Z_{h^*-1}}$, we prove in the next section
that the integration in the r.h.s. in \equ(uu) is well defined.
This will be proved by
using the following lemma (whose proof is similar to the proof of lemma
2.8).
\vskip.5cm
\0{\bf 2.10}{\rm LEMMA} {\it Assume that $h^*$ is finite uniformly in $L,\b$
so that $|{\s_{h^*-1}\over\g^{h^*}}|\ge \bar\k$, if $\bar\k$ is a constant.
Then it is possible to find a constant $C_N$ such that,
for $|x-y|\le{L\over 2}$, $|x_0-y_0|\le{\b\over 2}$
$$|g^{(\le h^*)}_{\o,\o'}(\xx;\yy)|\le {C_N (\max
\{\g^{h^*},L^{-1}\})\over
1+(\g^{h^*}|\xx-\yy|)^N}\Eq(2.50)$$}
\vskip.5cm
{\bf 2.11} From Lemma 2.7 we see that it is possible to have
quartic or bilinear contribution to $\VV^{(h)}$, if $|h|$ is large enough,
with $\sum_i\s_i\oo_i p_F+2 m p_L=0$ ${\rm mod.}\; 2\pi$
only with a extremely large $m$, namely $|m|\ge C\g^{-h\over\t}$;
to show that such terms are irrelevant we use the fact that
$|\hat\phi_m|\le Ce^{-\x |m|}$, see 3.6.
Comparing Lemma 2.8 and Lemma 2.10 we see that the propagator of the
integration of
all the scale between $h^*$ and $h_\b$ has the same bound as the
propagator
of the integration of a single scale greater than $h^*$; this will be
used to perform the integration of all the scales $\le h^*$ in a single step.
In fact $\g^{h^*}$ is a momentum scale and, roughly speaking,
for momenta bigger than $\g^{h^*}$ the theory is "essentially"
a massless theory (up to $O(\s_h\g^{-h})$ terms) while for momenta
smaller than $\g^{h^*}$ is a "massive" theory with mass $O(\g^{h^*})$.

\vskip2.cm
\centerline{\titolo 3. Analiticity of the effective potential}
\*\numsec=3\numfor=1
\0{\bf 3.1} We find  convenient in order to discuss the
convergence of the expansion for
the effective potential $\VV^{(h)}$ to pass to the
coordinate representation.

Let be
$$\psi^{(\le h)\sigma}_{\xx,\o}={1\over L\b}\sum_{\kk'\in {\cal D}_{L,\b}}
e^{i\sigma\kk'\xx}
\psi^{(\le h)\sigma}_{\kk'+\o \pp_F,\o}$$
and by \equ(loc1) \equ(loc2) we have, if $\RR=1-\LL$:
\vskip.5cm
1)for $n>4$
$$\sum_{x_1,...,x_n\in\L}\int_{-\b/2}^{\b/2}dx_{0,1}...dx_{0,n}
\RR\{\prod_{i=1}^n\psi^{(\le
h)\sigma_i}_{\xx_i,\o_i}
W^{h}_{n,m}\}=$$
$$\sum_{x_1,...,x_n\in\L}\int_{-\b/2}^{\b/2} dx_{0,1}...dx_{0,n} \prod_{i=1}^n\psi^{(\le
h)\sigma_i}_{\xx_i,\o_i}
W^{h}_{n,m}$$
\vskip.5cm
2)for $n=4$ and $\sum_{i=1}^4\s_i\omega_i p_F+2mp_L\not=0$ ${\rm
mod.}\; 2\pi$ one has $\RR=I$
otherwise
$$\sum_{x_1,...,x_4\in\L}\int_{-\b/2}^{\b/2}
dx_{0,1}...dx_{0,4}\prod_{i=1}^4\psi^{(\le
h)\sigma_i}_{\xx_i,\o_i}
\RR\{W_{4,m}^{h}(\xx_1-\xx_4,\xx_2-\xx_4,\xx_3-\xx_4;x_4;\{\o\})\}= \;
\Eq(za.1)  $$
$$ \sum_{x_1,...,x_4\in\L}\int_{-\b/2}^{\b/2}dx_{0,1}...dx_{0,4}
\prod_{i=1}^4\psi^{(\le h)\sigma_i}_{\xx_i,\o_i}
[ W_{4,m}^{h}(\xx_1-\xx_4,\xx_2-\xx_4,\xx_3-\xx_4;x_4;\{\o\})-$$
$$-\d(x_{0,1}-x_{0,2})\d(x_{0,2}-x_{0,3})\d(x_{0,3}-x_{0,4})\d_{x_1,x_2}
\d_{x_2,x_3}\d_{x_3,x_4}
\sum_{{\bf t}_1,{\bf t}_2,{\bf t}_3\in\L}
W_{4,m}^{h}({\bf t}_1,{\bf t}_2,{\bf t}_3;\{\o\})]$$
where we have used that if $\sum_{i=1}^4\s_i\omega_i p_F+2mp_L=0$ ${\rm
mod.}\; 2\pi$ the kernels
$W^{h}_{4,m}$ are {\it translation invariant}.
\vskip.5cm
3)for $n=2$ and $\sum_{i=1}^2\s_i\omega_i p_F+2mp_L\not=0$ ${\rm
mod.}\; 2\pi$
then $\RR=I$
otherwise
$$\sum_{x_1,x_2\in\L}\int_{-\b/2}^{\b/2}dx_{0,1}
dx_{0,2}\RR\{\prod_{i=1}^2\psi^{(\le h)\sigma_i}_{\xx_i,\o_i}
W_{2,m}^{h}(\xx_1-\xx_2;x_2;\{\o\})\}=$$
$$\sum_{x_1,x_2\in\L}\int_{-\b/2}^{\b/2} dx_{0,1}dx_{0,2}
\prod_{i=1}^2\psi^{(\le h)\sigma_i}_{\xx_i,\o_i}
[W^{h}_{2,m}(\xx_1-\xx_2;x_2;\{\o\})-$$
$$\d(x_{0,1}-x_{0,2})\d_{x_1,x_2}
\sum_{{\bf t}\in\L} W_{2,m}^{h}({\bf t};\{\o\})+\; \Eq(za.2)$$
$$\partial_{x_2^0} \d(x_{0,1}-x_{0,2})\d_{x_1,x_2}
\sum_{{\bf t}\in\L} t_0 W_{2,m}^{h}({\bf t};\{\o\})+
\d(x_{0,1}-x_{0,2})\partial_{x_2}\d_{x_1,x_2}
\sum_{{\bf t}\in\L} t W_{2,m}^{h}({\bf t};\{\o\})+$$
$$\o_1\d(x_{0,1}-x_{0,2})\d^2_{x_1,x_2}
\sum_{{\bf t}\in\L} t W_{2,m}^{h}({\bf t};\{\o\})]$$
and again we have used that when the $\RR$ operation acts
non trivially the kernels are
translation invariant; moreover $\partial_{x_2}\d_{x_1,x_2}={1\over
2}[\d_{x_1,x_2+1}-\d_{x_1,x_2-1}]$ and $\partial^2_{x_2}\d_{x_1,x_2}=
{i\cos p_F\over 2}[\d_{x_1,x_2+1}+\d_{x_1,x_2-1}-2\d_{x_1,x_2}]$.
\vskip.5cm
The following identities, obtained by \equ(za.1) \equ(za.2),
will be useful in the
following
$$\sum_{\xx_1,...,\xx_4\in\L}\int_{-\b\over 2}^{\b\over 2}
dx_{0,1}...dx_{0,4}(\xx_i-\xx_j)^a
\RR\{\prod_{i=1}^4\psi
^{(\le h)\sigma_i}_{\xx_i,\o_i}
W_{4,m}^{h}(\xx_1-\xx_4,\xx_2-\xx_4,\xx_3-\xx_4;x_4;\{\o\})\}=\Eq(za.3)$$
$$\sum_{x_1,...,x_4\in\L}\int_{\b\over 2}^{\b\over
2}dx_{0,1}...dx_{0,4}(\xx_i-\xx_j)^a
\prod_{i=1}^4\psi^{(\le h)\sigma_i}_{\xx_i,\o_i}
W_{4,m}^{h}(\xx_1-\xx_4,\xx_2-\xx_4,\xx_3-\xx_4;x_4;\{\o\})$$
if $a$ is any integer $\geq 1$ and $i,j$ can take any value between
$1$ and $4$.
\vskip.5cm
$$\sum_{x_1,x_2\in\L}\int_{-\b\over 2}^{\b\over 2} dx_{0,1}
dx_{0,2}(x_1-x_2)^a\RR\{
\prod_{i=1}^2\psi^{(\le h)\sigma_i}_{\xx_i,\o_i}
W_{2,m}^{h}(\xx_1-\xx_2;x_2;\{\o\})\}=\;  $$
$$\sum_{x_1,x_2\in\L}\int_{-\b\over 2}^{\b\over 2} dx_{0,1}dx_{0,2}[
\prod_{i=1}^2\psi^{(\le
h)\sigma_i}_{\xx_i,\o_i}(x_1-x_2)^a
W_{2,}^h(\xx_1-\xx_2;x_2;\{\o\})-$$
$${1\over 2}\psi^{\s_1(\le h)}_{\xx_1}
[\psi^{\s_2(\le h)}_{x_1+1,x_{0,1}}(-1)^a-\psi^{\s_2(\le h)}_{x_1-1,x_{0,1}}]
(x_1-x_2) W^h_{2,m}(\xx_1-\xx_2;x_2;\{\oo\})+\Eq(zaa.3)$$
$$+{i\o\over 2}\cos(p_F)\psi^{\s_1(\le h)}_{\xx_1}
[\psi^{\s_2(\le h)}_{x_1-1,x_{0,1}}+(-1)^a\psi^{\s_2(\le h)}_{x_1+1,x_{0,1}}]
(x_1-x_2) W^h_{2,m}(\xx_1-\xx_2;x_2;\{\oo\})]$$
if $a\ge 1$ and a similar equation hold if $(x_1-x_2)$ is replaced by
$(x_{0,1}-x_{0,2})$.
\vskip.5cm

Of course one can integrate the $\d$ functions in \equ(za.1) \equ(za.2) so
obtaining
$$\sum_{x_1,...,_4\in\L}\int_{-\b\over 2}^{\b\over 2}dx_{0,1}...dx_{0,4}
\RR\{\prod_{i=1}^4\psi^{(\le h)\sigma_i}_{\xx_1,\o_i}
W_{4,m}^{h}(\xx_1-\xx_4,\xx_2-\xx_4,\xx_3-\xx_4;x_4;\{\o\})\}=\;  \Eq(z.7)$$
$$\sum_{x_1,...,_4\in\L}\int_{-\b\over 2}^{\b\over 2}W_{4,m}^{
h}(\xx_1-\xx_2,\xx_2-\xx_4,\xx_3-\xx_4;x_4;\{\o\})$$
$$[\psi^{(\le
h)\sigma_1}_{\xx_1,\o_1}\psi^{(\le
h)\sigma_2}_{\xx_2,\o_2}\psi^{(\le h)\sigma_3}_{\xx_3,\o_3}\psi^{(\le
h)\sigma_4}_{\xx_4,\o_4}-\psi^{(\le
h)\sigma_1}_{\xx_1,\o_1}\psi^{(\le
h)\sigma_2}_{\xx_1,\o_2}\psi^{(\le h)\sigma_3}_{\xx_1,\o_3}\psi^{(\le
h)\sigma_4}_{\xx_1,\o_4}]$$
The square brakets in the above equation can be written as
$$\psi^{(\le h)\sigma_1}_{\xx_1,\o_1} D^{(\le
h)\sigma_2}_{\xx_{2,1},\o_2}\psi^{(\le h)\sigma_2}_{\xx_3,\o_3}\psi^{(\le
h)\sigma_4}_{\xx_4,\o_4}$$
$$+\psi^{(\le h)\sigma_1}_{\xx_1,\o_1}\psi^{(\le h)\sigma_2}_{\xx_1,\o_2}
D^{(\le h)\sigma_3}_{\xx_{3,1},\o_3}\psi^{(\le h)\sigma_4}_{\xx_4,\o_4}+
\psi^{(\le h)\sigma_1}_{\xx_1,\o_1}\psi^{(\le
h)\sigma_2}_{\xx_1,\o_2}\psi^{(\le h)\sigma_3}_{\xx_1,\o_3} D^{(\le
h)\sigma_4}_{\xx_{1,4},\o_4}$$
so that the effect of $\RR$ is essentially to change the coordinate
of the $\psi^{(\le h)\sigma_i}_{\xx_i,\o}$ fields and to replace
a field $\psi^{(\le h)\sigma_i}_{\xx_i,\o}$
with $D^{(\le h)\sigma_i}_{\xx_{ij}}$ where
$$D_{\xx_{ij}}^{(\le
h)\sigma_i}=\psi^{(\le
h)\sigma_i}_{\xx_i,\o_i}-\psi^{(\le h)\sigma_i}_{\xx_j,\o_i}=
(\xx_i-\xx_j)\int_0^1 dt\vec\partial
\psi^{(\le h)\sigma_i}_{\xx_{ji}(t),\o_i}\; , \Eq(za.8)$$
with $\xx_{ji}(t)=t \xx_j+(1-t) \xx_i$ and $\vec\partial=(\partial_x,\partial_{x_0})$.
In the
same way integrating the $\d$'s in \equ(za.2) one can see
that the effect of $\RR$ is to replace a field $\psi^{(\le
h)\sigma}$
with a field $A^{(\le
h)\sigma}$ which can be written as:
$$(\xx_2-\xx_1)^2\int_0^1 dt_1 \int_{0}^{t_1}
dt_2\partial^2\psi^{(\le
h)\sigma}_{x_{21}(t_2)}+(x_2-x_1)\D\psi_{x_2}^{(\le
h)\sigma}\;  \Eq(za.9)$$
where $\D\psi_{x_2}^{(\le
h)\sigma}$ is defined implicitely by the above equation and it
has the same scaling properties of $\partial^2\psi_{x_2}^{(\le
h)\sigma}$.
The $t$-parameters appearing in \equ(za.8)\equ(za.9) are called
{\it interpolation parameters}.
\vskip.5cm
\0{\bf 3.2} From \equ(z.7),(3.6) one can see that
the effect of $\RR$ is that a derivative is applied
on a field and there is a factor $(\xx_i-\xx_j)$ more in the kernels.
With respect to the $\RR=1$ case the presence of the derivative
produce a gain in the bounds and the factor
$(\xx_i-\xx_j)$ a loss (we are studying the large distance behaviour)
and at the end, due to the structure of the iterative integration
discussed in sec.2, the final effect is a gain. This is standard in
renormalization group formalism, see [G2]. Analogous considerations
can be done for \equ(za.9).

From \equ(za.3) we see that, by definition, $\RR$ does not act
if there is a factor $(\xx_i-\xx_j)^a$ in front of $\RR W^h_{4,m}$.
This property will simplify
the discussion and was already used
in [BM1], if the fermions are on the continuum and not on a chain;
\equ(zaa.3) is the analogue for the $n=2$ case.
Note the two different equivalent ways to write the $\RR$ operation:
in \equ(za.1),\equ(za.2) the $\RR$ acts on the kernels
of the effective potential; we can say that
it consists in replacing $W^h_{n,m}$ with $\RR W^h_{n,m}$
where $\RR W^h_{n,m}$ are the terms in square brakets in
r.h.s of \equ(za.1),\equ(za.2), leaving the fermionic fields unchanged;
on the other hand writing as in \equ(z.7)
the $\RR$ operation acts on the $\psi$ fields, leaving the kernel
$W^h_{n,m}$ unchanged.
\vskip.5cm
\0{\bf 3.3} From \equ(3.20h) we have that, if
$E^T_{h+1}$ denotes the {\it truncated expectation} with propagator
${g^{(h+1)}(\xx;\yy)\over Z_h}$\equ(3.13a):
$$\VV^h(\sqrt{Z_h}\psi^{\le h})=\sum_{n=1}^{\io}{1\over n!}
(-1)^{n+1}E^T_{h+1}(\hat
\VV^{h+1}
(\sqrt{Z_h}\psi^{\le h+1}),...)$$
where $\hat \VV^{h+1}$ is obtained from $\VV^{h+1}$ following the
operations described in sec.2.
Iterating the above equation we obtain that
the effective potential can be written in term of a {\it
tree expansion} [G2].
$$   $$
$$   $$
$$   $$
$$   $$
$$   $$
$$   $$
$$   $$
A tree $\t$ consists of a family of lines arranged to connect a partially
ordered set of points ({\it vertices}). Every line has two vertices at the
extreme except the first one which has only one vertex, the first
vertex $v_0$. The other extreme $r$ is the {\it root} and it is not a
vertex. From each vertex $v$ start $s_v$ lines;
if $s_v=0$ the vertex is called {\it end-point},
if $s_v=1$ is called {\it trivial vertex} and if $s_v>1$ is called {\it non
trivial vertex}. To each vertex $v$ we associate a {\it scale} $h_v$.
The vertices of $\t$ are naturally (partially)
ordered. Giving to the tree an orientation from left to
right we write $v_1<v_2$ (so that $h_{v_1}<h_{v_2}$)
if $v_1$ is before
$v_2$
on the tree. Two trees that can be superposed by a suitable continuous
deformation are equivalent \ie the trees are topological trees. The number
of trees with
$n$ end-points is bounded by $C^n$, if $C$ is a constant.

To each end point $v$ with scale
$h_v$ we associate either
one of the addend of $\LL V^{(h_v-1)}$ in \equ(3.11ah)
or one of the
terms of $\RR V^0$; this second case is possible only if the scale
of the end point is $1$. For simplicity of notations
we assume in this section
$$V^0=\sum_{m\not =0,\pm
\bar m}e^{i2mp_L x}\sum_{x\in\L}\int_{-\b/2}^{\b\over2}dx_0
\hat\phi_m \psi^{(\le
h)+}_\xx \psi^{(\le h)-}_\xx\Eq(R)$$
It will be clear from the following consideration that the general case
is completely equivalent.
Between two non trivial vertices $v_1,v_2$ there are all
the trivial vertices $v$ with scales $h_{v_1}< h_v<h_{v_2}$.
If $n\geq 2$ and the scale of an end point
is different from $1$ there is no trivial vertex
between the
end-point and the non trivial
vertex $v$ immediately preceding it
on the tree, so that the scale of the end point is $h_v+1$ and the running
coupling constants associated to it is one of the $\vec v_{h_v}$.
If $n=1$ and to
the end point is associated a running coupling constant
there is only a vertex $v_0$ on the tree, besides the end-point, and the scale
of the running coupling constant is $h_{v_0}=h_{k+1}$. Each tree $\t$ take a label
distinguishing which term is associated to the end points.
Each trivial or non trivial
vertex carry a label $\RR$ except $v_0$ which can carry
an $\RR$ or $\LL$ operation. The set of all the trees
with $n$ end points with root with scale $k$ will be denoted
by $\t_{n,k}$.

A {\it cluster } $L_v$ with frequeny $h_v$ is
the set of the end points (possibly only one)
reachable from the vertex $v$,
and the tree provides an organization of end points into a hierarchy of clusters.
Given a cluster $L_v$ (with scale $h_v$), we define
$$N_v=\sum_{i\in L_v} m_i\Eq(n)$$
where the index $i$ denotes the end-points contained in $L_v$,
$m_i=0$ if to the end-point $i$ is associated a running coupling constant
not of the kind $\t,\iota$, $m_i=\pm \bar m$ if
to the end-point $i$ is associated a running coupling constant
of the kind $\t,\iota$ and finally
if to the end point $i$ is associated
one of the irrelevant terms \equ(R) $m_i\not= 0,\pm \bar m$.
By definition
$$N_v=N_{v^1}+...+N_{v^{s_v}}\Eq(non)$$
and if $v_i$ is an end point than $N_v=m_i$.

Following standard arguments (see [G2]) the effective potential can
be written in the following way

$$\VV^{(k)}(\sqrt{Z_k}\psi^{(\le k)})=\sum_{n=1}^\io\sum_{\t\in\t_{k,n}}
V^k(\t,\sqrt{Z_k}\psi^{(\le k)})\Eq(z.10)$$
If $v_0$ is the first vertex after the root, $\t_1,..,\t_{s_{v_0}}$ are
the subtrees
starting after the vertex $v_0$, then $V^k(\t,\sqrt{Z_k}\psi^{(\le k)})$ is defined
inductively
by the relation
$$V^k(\t,\sqrt{Z_k}\psi^{(\le k)})={1\over s_{v_0}!} E^T_{k+1}[\bar
V^{k+1}(\t_1,\sqrt{Z_{k}}\psi^{(\le k+1)}),.., \bar
V^{k+1}(\t_{s_{v_0}},\sqrt{Z_{k}}\psi^{(\le k+1)})]\Eq(z11)$$
where $\bar
V^{k+1}(\t_i,\sqrt{Z_{k}}\psi^{(\le k+1)})$
\vskip.5cm
1)is equal to $\RR
\hat V^{k+1}(\t_i,\sqrt{Z_{k}}\psi^{(\le k+1)})$ if the subtree $\t_i$
is not trivial \ie if the first vertex of $\t_i$ is not an and-point,
\vskip.5cm
2)if
the first vertex of $\t_i$ is an end-point $\bar
V^{k+1}(\t_i,\sqrt{Z_{k}}\psi^{(\le k+1)})$
is equal to one addend
of $\LL \hat V^{k+1}(\t_i,\sqrt{Z_{k}}\psi^{\le k+1})$ \equ(3.11ah)
or, if $k=-1$, to one term of
$V^{0}(\t_i,\psi^{\le k+1})$.
\vskip.5cm
Let we assume that the effective potential can be written as:

$$V^k(\t,\sqrt{Z_k}\psi^{(\le k)})=\int d\xx_{v_0}\sum_{P_{v_0}}
\sqrt{Z_k}^{|P_{v_0}|}
\tilde\psi^{(\le k)}(P_{v_0}) W^k(\t,P_{v_0},\xx_{v_0})\Eq(zz10)$$
where $\int
d\xx_{v_0}=\prod_{f\in I_{v_0}}\sum_{x(f)\in\L}\int_{-\b/2}^{\b/2}
dx(f)$,
$P_{v_0}$ is a non empty set of $I_{v_0}$, the
field labels associated with the end-points
reachable from $v_0$ (\ie all of them), $|P_{v_0}|$ is the number of
elements of $P_{v_0}$, $\sum_{P_{v_0}}$ is the sum over such sets,
$x_{v_0}$ is the set $\{ x(f)\}$, $f\in I_{v_0}$ \ie the set
of the coordinates associated with the tree and:
$$\tilde\psi^{(\le k)}(P_{v_0})=
\prod_{f\in P_{v}}\psi^{(\le k)\s(f)}_{\xx(f),\o(f)}\Eq(3.13a)$$

The above assumption is proved by
induction assuming that \equ(zz10) holds also for the subtrees $\t_i$
and using \equ(z11).
In fact by
the identity
$$\tilde\psi^{(\le k+1)}(P)=\sum_{Q\subset P}
\tilde\psi^{(< k+1)}(Q)  \tilde\psi^{(k+1)}(P/Q)$$
we obtain
$$V^k(\t,\sqrt{Z_k}\psi^{(\le k)})={1\over s_{v_0}!}
\sum_{P_{v_0}}\sqrt{Z_{k}}^{|P_{v_0}|}
\sum_{P_{v_0^1},..,P_{v_0^{s_{v_0}}} }
\sum_{ Q_{v_0^1},..,Q_{v_0^{s_{v_0}}} } $$
$$\E_{k+1}^T [\int d\xx_{v_0^1} \tilde\psi^{(k+1)}(P_{v_0^1}/Q_{v_0^1})
\bar W^{k+1}(\t_1,P_{v_0^1},\xx_{v_0^1}),...,$$
$$\int
d\xx_{v_0^{s_{v_0}}}
\tilde\psi^{(k+1)}(P_{v_0^{s_{v_0}}}/Q_{v_0^{s_{v_0}}})\bar
W^{k+1}(\t_{s_{v_0}},P_{v_0^{s_{v_0}}} ,\xx_{v_0^{s_{v_0}}})]
\tilde\psi^{(\le k)}(P_{v_0})$$
where $\E_h^T$ denotes the truncated
expectation with respect to the propagator $g^{(h)}(\xx;\yy)$
\equ(3.15a),
$P_{v_0}\cup_i Q_{v_0^i}$, $Q_{v_0^i}\subset P_{v_0^i}$,
$v_{0}^i$ is the first
vertex of the subtree $\t_i$, and
$\bar
W^{k+1}(\t,P_{v_0^i},\xx_{v_0^i})=\RR
\bar W^{k+1}(\t,P_{v_0^i},\xx_{v_0^i})$
if $\t_i$ is not an end-point,
$\bar
W^{k+1}(\t,P_{v_0^i},\xx_{v_0^i})=\vec v_{k+1}$ if it is an end-point but $k+1\not=0$
and if it is an end-point with $k+1=0$ is equal to $\vec v_0$ or to
$\hat\phi_m e^{i2mp_L x}$.
The above expression of course proves \equ(zz10).

We prove the following theorem, if $\vec v_h$ are the running coupling constants in
\equ(3.11ah)
\vskip.5cm
\0{\bf 3.4}\0{\cs Theorem} {\it Let be $k> h^*$ given by \equ(h*).
There exists a constant $\bar\e_k$ such that, if
$\sup_{h>k}|\vec v_h|\le\bar\e_k$ and
$\sup_{h>k}|{Z_h\over Z_{h-1}}|\le e^{c_1\bar\e_k^2}$,
chosen $\g^{1\over\t}/2\ge 1$ and $L,\b\ge \g^{-h^*}$,
then
$$|\int d\xx^{P_{v_0}}\sum_{\t\in\t_{n,k}}|W^k(\t,P_{v_0},\xx^{P_{v_0}})|
(1+\g^k d(P_{v_0})^N)|\le L\b \g^{-kD(P_{v_0})}(C_N\bar\e_k)^n
e^{-\g^{-k/\t} c_2 s_{P_v}\over 2^{-k+3}}$$
where $s_{P_{v_0}}=1$ if $N_{v_0}\not=0$, $|P_{v_0}|=2$ and $0$ otherwise,
$N$ is a positive integer, $\xx^{P_{v^0}}$ is the set of points associated to
$P_{v_0}$,
$c_1,c_2,C_N$ are constants (not depending on $L,\b$), $d(P_{v_0})$ is the length
of the shortest tree connecting the set of points $\xx^{P_{v_0}}$,
and
$$D(P_{v_0})=-2+\sum_{f\in P_{v_0}}(1/2+m_f)$$
where $m_f$ is the order of the derivative applied to the fields of label
$f$.}
\vskip.5cm
\0{\bf 3.5} {\cs PROOF} In order to write in an explicit form \equ(z11) it is convenient
to start studying $\RR \hat V^h$:
$$\RR \hat V^k(\t,\sqrt{Z_{k-1}}\psi^{(\le k)})={1\over
s_{v_0}!}\sum_{P_{v_0}}(\sqrt{Z_{k-1}})^{|P_{v_0}|}
\RR\{ \sum_{ P_{v_0^1},..,P_{v_{0^s_{v_0}}} }
\sqrt{Z_{k}\over Z_{k-1}}^{|P_{v_0}|}$$
$$\sum_{ Q_{v_0^1},..,Q_{v_0^{s_{v_0}}} }
\E_{k+1}^T[\int d\xx_{v_0^1} \tilde\psi^{(k+1)}(P_{v_0^1}
\setminus Q_{v_0^1})\bar
W^{k+1}(\t,P_{v_0^1},\xx_{v_0^1}),...,$$
$$\int d\xx_{v_0^{s_{v_0}}}
\tilde\psi^{(k+1)}(P_{v_0^{s_{v_0}}} \setminus Q_{v_0^{s_{v_0}}})\bar
W^{k+1}(\t,P_{v_0^{s_{v_0}}},\xx_{v_0^{s_{v_0}}})]
\tilde\psi^{(\le k)}(P_{v_0})\}\Eq(bab)$$

From \equ(z.7)-\equ(za.9) we know that
the effect of $\RR$ is simply to replace $\tilde\psi^{(\le k)}(P_{v_0})$ with
$$\sum^*_{i,j}(\xx_i-\xx_j)^{a_{v_0}}
\hat\psi^{\le k}(P_{v_0})\Eq(bab1)$$
(in the following $\sum^*_{i,j}$ will be omitted), where
$\xx_i,\xx_j$ are two coordinates in $\xx^{P_{v_0}}$ and
\vskip.5cm
1)$a_{v_0}=0$ and $\hat\psi^{(\le k)}(P_{v_0})=\tilde\psi^{(\le
k)}(P_{v_0})$
if $|P_{v_0}|>4$ or $\sum_{f\in P_{v_0}}\s(f)\oo(f) p_F+2 N_{v_0} p_L\not=0$
${\rm mod.}\; 2\p$,
see \equ(loc1),\equ(loc2).
\vskip.5cm
2)if $|P_{v_0}|=4$ and
$\sum_{f\in P_{v_0}}\s(f)\oo(f) p_F+2 N_{v_0} p_L=0$
${\rm mod.}\; 2\pi$ then
$a_{v_0}=1$
and $\hat\psi^{(\le k)}(P_{v_0})$ differs from $\tilde\psi^{(\le
k)}(P_{v_0})$
because a field $\int_0^1 dt \partial\psi^{(\le k)}_{x_{ij}(t)}$
replaces a $\psi^{(\le k)}$ field,
and the remaning $\psi$ fields are applied on different coordinates among
$\xx^{P_{v_0}}$, see \equ(z.7);
\vskip.5cm
3)if $|P_{v_0}|=2$ and
$\sum_{f\in P_{v_0}}\s(f)\oo(f) p_F+2 N_{v_0} p_L=0$
${\rm mod.}\; 2\pi$ then $a_{v_0}=2$
and $\hat\psi^{(\le k)}(P_{v_0})$ differs from
$\tilde\psi^{(\le k)}(P_{v_0})$
because a field $\int_0^1 dt_1 \int_0^{t_1} dt_2
\partial^2\psi^{(\le k)}_{x_{ij}(t_2)}$
or a ${\D\psi^{(\le k)}\over (x_1-x_2)}$
replaces a $\psi^{(\le k)}$ field.
\vskip.5cm
Let we write:
$$\hat\psi^{(h)}(P)=\prod_{f\in
P}\partial^{q(f)}_{\xx(f)}\psi_{\xx(f)}^{\s(f)(h)}\Eq(bab7)$$
where $q=0,1,2,3$, $\partial^0=1$, $\partial^1=\partial_\xx$,
$\partial^2_\xx=\partial_\xx
\partial_\xx$,
$\partial^3={1\over (x_1-x_2)}\D$.

We remember the well known expansion of truncated expectation in term of interpolating parameters
$s_t$, $t=1,..,k-1$:
$$\E^T_{h}(\hat\psi^{(h)}(P_1),...,\hat\psi^{(h)}(P_k))=
\sum_{T}\prod_{l\in
T}
\partial^{q_l}_{\xx_l} \partial^{q'_l}_{\yy_l} g^{(h)}(\xx_l;\yy_l)\int
dP_{T}(s) det G^{h,T}(s)\Eq(bab2)$$

where $T$ is a set of lines forming an {\it anchored tree graph} between the
clusters of vertices from which the fields labeled with $P_1,..,P_k$
emerge: this means that $T$ is a set of lines connecting two points
in different clusters, which becomes a tree graph if one identifies
all the points in the same cluster; if $l\in T$ $\xx_l,\yy_l$
are the end-points of the line and are
such that $\xx_l=\xx_{ij}$ or $\yy_l=\xx_{i'j'}$, where
$\xx_{ij}$ denotes the coordinate of the $i$-th field
of the monomial $\tilde\psi(P_j)$. If $f$ is a field variable such that
$\xx(f)\equiv \xx_l\equiv\xx_{ij} $, we write $q(f)\equiv q_l\equiv q_{ij}$.
In the same way if $\bar f$ is such that $\xx(\bar f)\equiv \yy_l\equiv\xx_{i'j'} $
we say  $q(\bar f)\equiv q'_l$.
$G^{h,T}(s)$ is a
$(n-k+1)
\times (n-k+1)$ matrix (if $n$ is the total number of fields), whose elements are
$G^T_{jij'i'}=S_{jj'} \partial^{q_{ij}}\partial^{q_{i'j'}}
g^h(x_{ij}-x_{i'j'})$ with $x_{ij}-x_{i'j'}$ non belonging
to $T$,
$S_{jj'}=\prod_{t=j}^{j'-1} s_t$ and $dP_T(s)$ is a normalized measure which
depends on $s_t$, $t=1,...,k-1$ and $T$ (for the well known
explicit formula of $dP_T$ and for its derivation, one can look for instance
[BGPS]).

We write in an explicit way the action of $\RR$ in \equ(bab) obtaining

$$\RR \hat V^k(\t,\sqrt{Z_{k-1}}\psi^{(\le k)})={1\over
s_{v_0}!}\sum_{P_{v_0}}
\sqrt{Z_{k-1}}^{|P_{v_0}|}\sum_{ P_{v_0^1},..,P_{v_0^{s_{v_0}}}}
\sqrt{Z_{k}\over Z_{k-1}}^{|P_{v_0}|}$$
$$\sum_{ Q_{v_0^1},..,Q_{v_0^{s_{v_0}}}} \int d\xx_{v_0}
[(\xx_i-\yy_j)^{a_{v_0}}
\E^T_{k+1}(\tilde\psi^{(k+1)}(P_{v_0^1} \setminus Q_{v_0^1}),...,
\tilde\psi^{(k+1)}(P_{v_0^{s_{v_0}}} \setminus Q_{v_0^{s_{v_0}}}) )
]\Eq(bab3)$$
$$\bar
W^{k+1}(\t_1,P_{v_0^1},\xx_{v_0^1})...\bar
W^{k+1}(\t_{s_{v_0}},P_{v_0^{s_{v_0}}},\xx_{v_0^{s_{v_0}}})
\hat\psi^{(\le k)}(P_{v_0})$$

We can write, for any anchored tree graph $T_{v_0}$ connecting the clusters
$L_{v_0^1}...L_{v_0^{s_{v_0}}}$:

$$(\xx_i-\yy_j)=\sum_{l\in T}(\xx_{l}-\yy_{l})+\sum^*_{iji'j'}(\xx_{ij}-\yy_{i'j'})\Eq(bab7)$$

where $\sum^*_{iji'j'}$ is a sum such that
there is no $l\in T_{v_0}$ such that
$\xx_{ij}=\xx_l,\yy_{i'j'}=\yy_l$.

So we can write:
$$(\xx_{i}-\yy_{i})^{a_{v_0}}\E^T_{k+1}(\tilde\psi^{(k+1)}(P_{v_0^1} 
\setminus Q_{v_0^1}),...
)
\bar
W^{k+1}(\t_1,P_{v_0^1},\xx_{v_0^1})...\bar
W^{k+1}(\t_{s_{v_0}},P_{v_0^{s_{v_0}}},\xx_{v_0^{s_{v_0}}})$$

$$=\sum_{T_{v_0}}\sum^*_{\{a\}}\{[\prod_{l\in T_{v_0}}
(\xx_l-\yy_l)^{a_l} g^{(k+1)}(\xx_l;\yy_l)]
[\int dP_{T_{v_)}}(s) det G^{k+1,T_{v_0}}(s)]$$

$$|\hat\xx_{v_0^1}|^{a_{v_0^1}}\bar
W^{k+1}(\t_1,P_{v_0^1},\xx_{v_0^1})...|\hat\xx_{v_0^{s_{v_0}}}|^{a_{v_0^{s_{v_0}}}}\bar
W^{k+1}(\t_{s_{v_0}},P_{v_0^{s_{v_0}}},\xx_{v_0^{s_{v_0}}})\}\Eq(zzz)$$
where $|\hat\xx_{v_0^i}|$ is the difference among two coordinates
in $\xx_{P^{v_0^i}}$, and $\sum^*_{\{a\}}$ is the sum over the indices
$a_l+a_{v_0^1}+..$ with the constraint that
$\sum_{l\in T_{v_0}}a_l+a_{v_0^1}+...+ a_{v_0^{s_{v_0}}}=a_{v_0}\le 2$.
We are now in position to iterate the above procedure studying each
$|\hat\xx_{v_0^i}|^{a_{v_0^i}}\bar
W^{k+1}(\t_i,P_{v_0^i},\xx_{v_0^i})$. We can distinguish several cases
\vskip.5cm
1)if $\t^i$ is a trivial tree, $a_{v_0^i}=0$ and
$|\hat\xx_{v_0^i}|^{a_{v_0^i}}\bar W^{k+1}=\vec v_{k+1}$;
\vskip.5cm
2)if $\t^i$ is not a trivial tree:
\vskip.5cm
2a)$a_{v_0^i}=0$. In this case we repeat word by word the
analysis for $W^k$ leading from \equ(bab) to \equ(bab3),
with the trivial substitution $k\to k+1$, $v_0\to v_0^i$. Note that if $\RR\not=0$
the effect of the renormalization is of replacing in the truncated expectation
in \equ(bab3) $\tilde\psi^{(k+1)}(P_{v_0^i} \setminus Q_{v_0^i} )$
with $\hat\psi^{(k+1)}(P_{v_0^i} \setminus Q_{v_0^i})$, defined in an
analogous way as
$\hat\psi(P_{v_0})$ \equ(bab1),
and to produce a factor $(\xx_i-\xx_j)^{a_{v_0^i}}$,
$\xx_i,\yy_j\in \xx^{P_{v_0^i}}$ which is written as in (3.20).
\vskip.5cm
2b)$a_{v_0^i}\not =0$. In this case from \equ(za.3)
we have $\RR=1$ if $|P_{v_0^i}|\ge 4$ and if $|P_{v_0^i}|=2$ we use
(3.4).
Also in this case we can repeat word by word the analysis for
$W^k$ leading from
\equ(bab) to \equ(bab3), with the trivial substitutions $k\to k+1,
v_0\to v_0^i$.
The crucial point is that the propagators belonging to
$T_{v_0^i}$ are of the form $(\xx_l-\yy_l)^{a_l} g(\xx_l;\yy_l)$ with $a_l\le
2$ \ie {\it the power $a$ is not increasing}.
In fact, in the case $|P_{v_0^i}|=4$ it could be {\it a priori}
a factor $(\xx_l-\yy_l)$ for the action of $\RR$ and another factor
$(\xx_l-\yy_l)$ from $|\hat\xx_{v_0^i}|^{a_{v_0^i}}$;
but $\RR=1$ from \equ(za.3) and so the first factor
is not present. Similar considerations
for the $|P_{v_0^i}|=2$ case.
\vskip1cm
Iterating this procedure
we find at the end
$$\RR\hat V^k(\t,\sqrt{Z_{k-1}}\psi^{(\le k)})=
\int d\xx_{v^0}
\sum_{P_{v^0}} \sqrt{Z_{k-1}}^{|P_{v_0}|}\tilde\psi^{(\le k)}(P_{v_0})
\sum_{\{ P_{v} \}}\{\int dt\}$$
$$[\prod_{\rm v\; not\; e.p.} |{Z_{h_v-1}
\over Z_{h_v-2}}|^{|P_v|/2}]
[\sum_{\{q\}}^* \prod_{\rm v\; not\; e.p.}{1\over s_v!}
\tilde\E^T_{h_v}(\hat\psi^{(h_v)}(P_{v^1} \setminus Q_{v^1}),...
,\hat\psi^{(h_v)}(P_{v^{s_v}} \setminus Q_{v^{s_v}})]$$
$$[\prod_{i\in V_1} (\vec v_{h_i})_a e^{2 i p_F x_i
\g_a}\g^{h_i\d_a}][\prod_{i\in V_2}
\hat\ph_{m_i}
e^{i2 m_i p_L\xx_i}]\Eq(pal)$$

where:
\vskip.5cm
1)The symbol $\sum_{\{ P_{v} \}}$ denotes the sum over all the compatible choices
of the sets $P_v$ on all the vertex of $\t$ except $v_0$; moreover
$Q_v\subset P_v$,
$P_v=\bigcup_i Q_{v^i}$.
\vskip.5cm
2)$\hat\psi^{(h_v)}(P_{v} \setminus Q_{v})=\prod_{f\in
P_{v} \setminus Q_{v}}\partial^{q_f}\psi^{\s(f)(h_v)}_{\xx_{ij}(t)}$
where $q_f=0,1,2,3$ with the same conventions as in \equ(bab7).
Moreover $\xx_{ij}(t)=\sum_{i'}\e_{i'j}(t) \xx_{i'j}$ with $\sum_{i'}\e_{i'j(t)}=1$
and $\e_{i'j}(0)=\d_{i,i'}$. Finally

$$\tilde\E^T_{h}(\hat\psi(P_1),...,\hat\psi(P_k))=\sum_{T_v}\sum^*_{\{a\}}
[\prod_{l\in T_v}
\partial^{q_l}_{\xx_l} \partial^{q'_l}_{\yy_l}
[(\xx_l-\yy_l)^{a_l}g^h(\xx_l;\yy_l)]]
\int dP_{T}(s) det {G}^T(s)$$

where $\sum^*_{\{q \}}$ and $\sum^*_{\{a \}}$
have the following constraints.
If $v$ is such that $|P_v|=4$ and the Kronecker $\d$ of \equ(loc1)
is verified
then $\hat\psi(P_v)$
contains a field $\partial\psi_{x'_{ij}}$ and 
there is a line $\bar l\in T_{\bar v}$,
$\bar v>v$, with $a_{\bar l}=1$ and for any $l\in T_{\hat v}$,
$l\not=\bar l$, $\bar v\ge \hat v>v$ $a_l=0$.
If $v$ is such that $|P_v|=2$ and the Kronecker $\d$ of \equ(loc2)
is verified then $\hat\psi(P_v)$
contains a field $\partial^2\psi_{\xx_{ij}}$ or $\D\psi$
and there are two (possibly coinciding) lines $l_1\in T_{\bar v_1}$, $l_2\in T_{\bar v_2}$
with
$\bar v_1>v$, $\bar v_2>v$
with $a_l=1$ and for any $l\in T_{\hat v}$, $l\not=l_1,l_2$  $\bar v_1\ge \hat v>v$,
$\bar v_2\ge \hat v>v$ one has $a_l=0$.
\vskip.5cm
4)$\{\int dt\}$ is a product of integral over the interpolation parameters.
\vskip.5cm
5) $V_1$ is
the set of the end points of the tree, and $V_2$ is the set of end points {\it not}
associated to running coupling constants.
Moreover $\g_a=\pm 1$ in correspondence of end-points of kind
$\t$ or $\iota$ and zero otherwise, and $\d_a=1$ in correspondence of end-points of kind
$\nu$ and zero otherwise.
\vskip.5cm
We sum \equ(pal) over the index $m_i$
\ie we sum over all the trees differing only
for the choices of the addend $\RR V^0$ associated to a given end-point(\ie
we sum over trees
with the same labels and differing only for the choices of $m_i$).
We can bound $det G^{h,T}$ by the well
known Gramm-Hadamard inequality
(see for instance [BGPS]).
Once that we have bounded the determinant, we have to make the integration
over the coordinates and over the interpolation variables.
It is convenient to change variables from $\{\xx\}$ to $\{{\bf r}\}$, where
$\{{\bf r}\}$ is the collection of the difference $\xx_l-\yy_l$
appearing in the factors
$\prod_l \partial^{q_l}_{\xx_l} \partial^{q'_l}_{\yy_l}
[(\xx_l-\yy_l)^{a_l} g^{h}(\xx_l;\yy_l)]$.
Note that $\xx_l\equiv x_{ij}(t)$ so that the determinant
of the Jacobian of this transformation is a function of $t$, and one can worry
about its integrability. However
it is possible to show (see [BM1], App. 3) that such determinant is exactly $1$.
So we obtain a bound for $W^k$ \equ(zz10),
estimating the propagators by lemma 2.8 (taking $L\ge \g^{-h^*}$ we get
$\max \{\g^h,L^{-1}\})=\g^{h}$),
using lemma 2.6 and repeating analogous considerations for $\LL \hat
V^k(\t,\sqrt{Z_{k-1}}\psi^{(\le k)})$:
$$L\b\g^{-k D(P_{v_0})}
\sum_{\{m_i\}} \sum_{\{P_v\}}[\prod_{\rm v\;
not\;e.p.}
\indic\big(|N_v|\ge
{\cal A}(\g^{-h_v\over\t}|P_v|^{-{1\over\t}}-\bar m |P_v|)\big)]
[\prod_{\rm v\; not\;e.p. } |{Z_{h_v-1}\over Z_{h_v-2}}|^{|P_v|\over
2}] $$
$$C^{|Q_v|-|P_v|} J(\t,P_{v_0})
[\prod_{\rm v\; not\; e.p.}\g^{-[D(P_v)+z_v(N_v,P_v)](h_v-h_{v'})}]
[\prod_{i\in V_1} (\vec v_{h_i})_a ][\prod_{i\in V_2}
|\hat\ph_{m_i}|]$$
where the factor
$\indic\big(|N_v|\ge
{\cal A}(\g^{-h_v\over\t}|P_v|^{-{1\over\t}}-\bar m |P_v|)\big)$
comes from lemma 2.7, as we know that the summand in \equ(pal)
is vanishing if the condition in the indicator function
is not verified,
$C$ is a constant and
$$J(\t,P_{v_0})=|
\prod_{\rm v\; not\; e.p.}{1\over s_v!}\sum_{T_v} 
\int dt \prod_{l\in T} \int d\{{\bf r}_l \g^{2 h_v}\}
\partial_{{\bf r_l}}^{q_l+q'_l}[
({\bf r})_l^a \g^{-h_v} [\g^{-h_v}\bar g^{h_v}({\bf r}_l)]|$$
and
$D(P_v)=-2+\sum_{f\in P_v}(1/2+m_f)$, with $m_f$ being the order of derivatives
applied to the field of label $f$, $v'$ is the vertex preceding $v$ on the tree (so that
$h_{v'}=h_v-1$). The presence $z_v(N_v,P_v)$
is due to the renormalization procedure and it is defined as
\vskip1cm
1)$z_v(N_v,P_v)=1$ if $|P_v|=4$,
$\sum_{f\in P_v} m_f=0$ and $\sum_{f\in P_v}\s(f)\o(f)p_F+2 N_v p_L=0$
\vskip.5cm
2)$z_v(N_v,P_v)=1$ if $|P_v|=2$, $\sum_{f\in P_v} m_f=1$
and $\sum_{f\in P_v}\s(f)\o(f)p_F+2 N_v p_L=0$
\vskip.5cm
3)$z_v(N_v,P_v)=2$ if $|P_v|=2$,
$\sum_{f\in P_v} m_f=0$ and $\sum_{f\in P_v}\s(f)\o(f)p_F+2 N_v p_L=0$
\vskip.5cm
Moreover $\int d\{{\bf
r}_l\}=\sum_{x_v\in\L}\int_{-\b}^{\b}\g^{\b}dr_{0,l}$

It is easy to check that $J(\t,P_{v_0})$ is bounded, if $\b,L\ge
\g^{-h^*}$, by
$\prod_{\rm v\;not\;e.p.} C^{\sum_i |P_{v^i}-|Q_{v^i}|}$
(the number of trees between the clusters $L_{v^1},..,L_{v_{s^v}}$
is just bounded by $s_v! C^{\sum_i |P_{v^i}|-|Q_{v^i}|})$.
\vskip.5cm
\0{\bf 3.6} We can write then
$$\prod_{\rm v\;not\;e.p. }
\g^{-[D(P_v)+z_v(N_v,P_v)](h_v-h_{v'})}\le$$
$$\prod_{\rm v\;not\; e.p. } \g^{-|P_v|\over 6}[\prod_{v\in
T_4}\g^{h_v-h_{v'}}]
[\prod_{v\in T_2}\g^{h_v-h_{v'}}][\prod_{v\in T_3}\g^{2(h_v-h_{v'})}]$$
where
\vskip1cm
1)$T_4$ is the set of vertices $v$
with $|P_v|=4$,
$\sum_{f\in P_v} m_f=0$ and $\sum_{f\in P_v}\s(f)\o(f)p_F+2 N_v p_L\not =0$
\vskip.5cm
2)$T_2$ is is the set of vertices $v$ with $|P_v|=2$,
$\sum_{f\in P_v} m_f=1$ and $\sum_{f\in P_v}\s(f)\o(f)p_F+2 N_v p_L\not =0$
\vskip.5cm
3)$T_3$ is is the set of vertices $v$ with $|P_v|=2$,
$\sum_{f\in P_v} m_f=0$ and $\sum_{f\in P_v}\s(f)\o(f)p_F+2 N_v p_L\not =0$.
\vskip.5cm
Let be $\TT$ the set of vertices $v$ (excluding end-points)
such that $v'$ is a non
trivial vertex; we say that $v$ is a {\it hard vertex}. To an
hard vertex $v$ we associate a
{\it depth} defined in the following
way:
if $v$ is the first hard vertex (\ie if $\bar v$ is any vertex on
the tree following $v$, $\bar v\not\in\TT$)
then $D_{v}=1$,
otherwise $D_{v}=1+max_{v''}\{D_{v''}\}$, where $v''$
are hard vertices following $v$ on the tree.
Note that  $D_{v}\le -h_{v'}+2$. We call $B_{D}$ the set of
$v\in \TT$ with depth $D$.

We can write
$$\prod_{\rm v\; not\; e.p.} \g^{-[D(P_v)+z_v(N_v,P_v)](h_v-h_{v'})}\le
\prod_{\rm v\; not\; e.p.} \g^{-|P_v|\over 6}[\prod_{v\in \TT_4}
\g^{-h_{v'}}]
[\prod_{v \in \TT_2}\g^{-h_{v'}}][\prod_{v\in \TT_3}
\g^{-2h_{v'})}]$$
where $v'$ is the vertex preceding $v$
and $\TT_i$ is the intersection
between $\TT$ and $T_i$.
It is easy to show that
$$\prod_{i\in V_2} |\hat\ph_{m_i}|\le e^{\x n\bar m\over 2}
\prod_{i\in V_2} e^{-\x |m_i|/2}\prod_{v\in \TT}
e^{-\x |N_{v}|\over 2^{D_{v}+1} }\Eq(fon)$$

In fact we write $\prod_{i\in V_2} |\hat\ph_{m_i}|\le
[\prod_{i\in V_2}
e^{-\x |m_i|/2}][\prod_{i\in V_2} e^{-\x |m_i|/2}]$ and we want to
show that
$$\prod_{i\in V_2} e^{-\x |m_i|/2}\le e^{\x \bar m n\over 2}
[\prod_{a=1}^{l}\prod_{v\in B_a}
e^{-\x |N_{v}|\over 2^{D_{v}+1} }]\Eq(fon1)$$
where $l\le -k+2$.
The l.h.s. of \equ(fon1) can be written as
$$[\prod_{i\in V_2\atop i\not\in L_v, v\in B_1} e^{-\x |m_i|/2}]
[\prod_{v_1\in B_1}\prod_{i\in L_{v_{1}}}e^{-\x |m_i|\over 2}]
\le$$
$$e^{\x \bar m n_1\over 2}[\prod_{i\in V_2\atop i\not\in L_v, v\in B_1}
e^{-\x |m_i|/2}]
[\prod_{v_1\in B_1} e^{-\x |N_{
v_{1}} |\over 4}][\prod_{v_1\in B_1}
e^{-\x |N_{v_{1}} |\over 4}]\Eq(fon2)$$
where $i\in L_{v_1}$ are the end-points in the cluster $L_{v_1}$
and we used \equ(non); $n_1$
are the end-point in $\cup_{v_1\in B_1 }L_{v_1}$
$v_1\in B_1$
of kinds $\iota,\t$. The r.h.s of \equ(fon2) can be bounded by
$$e^{\x \bar m n_1\over 2}
[\prod_{i\in V_2\atop
i\not\in L_v,  v\in B_1\cup B_2}e^{-\x |m_i|/2}]
[\prod_{v_1\in B_1\atop L_{v_1}\not\i L_{v_2}, v_2\in B_2 }
e^{-\x {|N_{v_{1}}|\over 4}}]$$
$$[\prod_{v_1\in B_1}
e^{-\x |N_{v_{1}} |\over 4}] \{ \prod_{v_2\in B_2} [
\prod_{v_1:L_{v_1}\i L_{v_2}}
e^{-\x |N_{v_{1}} |\over 4}] [\prod^*_{i\in L_{v_2}}
e^{-\x |m_i| \over 2}]\}\le$$
$$e^{\x \bar m(n_1+n_2)\over 2}
[\prod_{\in V_2\atop
i\not\in L_v,  v\in B_1\cup B_2}e^{-\x |m_i|/2}]
[\prod_{v_1\in B_1\atop L_{v_1}\not\i L_{v_2}, v_2\in B_2 }
e^{-{\x |N_{v_{1}}|\over 4}}]$$
$$[\prod_{v_1\in B_1}
e^{-\x |N_{v_{1}} |\over 4}] [\prod_{v_2\in B_2}e^{-\x
|N_{v_{2}}|\over 8}][\prod_{v_2\in B_2}e^{-\x
|N_{v_{2}}|\over 8}]$$
where $n_2$ are the end-points
in the clusters $\cup_{v_2\in B_2}L_{v_2}$
but not in any smaller ones contained in $v_2$
of kinds $\iota,\t$, and $\prod^*_{i\in L_v}$
is the product over the end-points in $L_v$ not contained in any
smaller cluster contained in $L_v$.
Proceding in this way we have \equ(fon1).

Using that $D_{v}\le -h_{v'}+2$ we have that, if
${\cal C}=\TT_2\bigcup\TT_3\bigcup\TT_4\i B$ and using \equ(fon1)
$$\prod_{\rm v\;not\;e.p.}
\indic\big(|N_v|\ge
{\cal A}(\g^{-h_v\over\t}|P_v|^{-{1\over\t}}-\bar m |P_v|)\big)
\prod_{i\in V_2} |\hat \ph_{m_i}|\le C^n
\prod_{i\in V_2} e^{-\x|m_i|/2}
\prod_{v\in {\cal C}}
e^{-\x C_2 \g^{-h_{v'}\over \t}\over 2^{-h_{v'}+3}}\Eq(sp)$$
where the constraint in the indicator function is used only for the
$v\in {\cal C}$, and $|P_v|\le 4$ for any $v\in {\cal C}$.

Then
$$\sum_{\{m_i\}}|\hat V^k(\hat\t,\sqrt{Z_{k-1}}\psi^{\le k})|\le
C_1^n\e^n \g^{-k D(P_{v_0})}\sum_{\{P_v\}}\prod_{\rm v\;not\;e.
p.}\g^{-|P_v|\over 8}$$
$$\prod_{v\in\TT_1}\g^{-h_{v'}}
e^{-\x C_2 \g^{-h_{v'}\over \t}\over 2^{-h_{v'}+3}}
\prod_{v\in\TT_2}\g^{-h_{v'}}e^{-\x C_2 \g^{-h_{v'}
\over \t}\over 2^{-h_{v'}+3}}
\prod_{v\in\TT_3}\g^{-2h_{v'}}
e^{-\x C_2 \g^{-h_{v'}\over \t} \over 2^{-h_{v'}+3}}$$

Choosing $\g$ so that $\g^{1\over\t}/2>1$
and remembering that the hard vertices are $\le C_1 n$
we have that, for $i=1,2,3$ ($C_1,C_2,C_3$ are constants)
$$\prod_{v\in\TT_i}\g^{-2h_{v'}}
e^{-\x C_2 \g^{-h_{v'}\over \t}\over 2^{-h_{v'}+3}}
\le C_2^n$$
By a standard calculation
$$\sum_{\t\in\t_{k,n}}\sum_{\{ P_v\}}\prod_{\rm v\;not\; e.p.}
\g^{-|P_v|\over 8}\le C_3^n$$
and this completes the proof of Theorem 3.3.
\vskip.5cm
{\bf Remark:} Formula \equ(sp) is the analogue of {\it Brjuno lemma} for
this problem; it ensures that the small denominator problem arising in the series for
the incommensurability
of $\phi_x$, can be controlled taking into account the Diophantine condition.
The same role is plaied by the original Brjuno lemma
in the proof of the convergence of the Lindstedt series.
The idea is quite simple: we can associate to each hard cluster
with four or two external lines an exponential factor
$e^{-\x |N_v| \over 2^{-h_{v'}+3}}$
which is indeed quite small if $|h_v|$ is big. It is due to the analyticity
of the incommensurate potential $\f_x$. But $|N_v|$ has to be very
large for the diophantine condition, as noted in lemma 2.7, and the resulting
factor compensates the "bad" factors $\g^{-h_{v'}}$ or $\g^{-2h_{v'}}$
due to the power counting. Note that the clusters which are not hard are
clusters in which the external lines are not contracted, or are contracted
with lines from the same cluster, so that it is sufficient to get
the factor $e^{-\x |N_v| \over 2^{-h_{v'}+3}}$ for the hard clusters.
The relevant or marginal terms in
\equ(loc1),\equ(loc2) are the analogue of the {\it resonances} in KAM
theory.
\vskip.5cm
We have now to perform the last integration \equ(uu).
\vskip.5cm
\0{\bf 3.7}{\rm THEOREM} {\it If there exists constants $\k$, $C_2$, $\e$
such that
${|\sigma_{h^*-1}|\over\g^{h^*}}\ge \k$
and $|\vec v_{h^*}|\le\bar\e_{h^*}$, $|{Z_{h^*}\over Z_{h^*-1}}|\le
e^{C_2\e_{h^*}^2}$ than
$$
|\int \tilde P_{Z_{h^*-1}}(d\psi^{(\le h^*)}) e^{-\tilde
\VV^{h^*}(\sqrt{Z_{h^*}}\psi^{(\le h^*)})}|\le
C_3 L\b \g^{2 h^*}\Eq(3.27)$$
where $C_3$ is a suitable constant}
\vskip.5cm
{\rm PROOF} The proof is rather straigthforward.
In fact we can write \equ(3.27) as a sum over trees, having only one non
trivial vertex $v$, $h_v\equiv h^*$, and $s_v$ end points. Then
\equ(3.27) can be written as
$\sum_{n=1}^\io\sum_{\t\in\t_n}V^{(h^*)}(\t)$,
if $\t_n$ are the trees with $n=s_v$ end points and $V^{h^*}(\t)$ is
given
by
$${1\over s_v!}E^T_{\le h^*}[\int d\xx_{v^1}\tilde \psi^{(\le
h^*)}(P_{v^1}) W^{h^*}(P_{v^1},\xx_{v^1})...\int d\xx_{v^{s_v}}
\tilde\psi^{\le h^*}(P_{v^{s_v}}) W^{h^*}(P_{v^{s_v}},\xx_{v^{s_v}})]$$
where $E^T_{\le h^*}$ denotes the integration with respect to the
propagator \equ(2.50) and $W^{h^*}(\xx_{v^i})=\sum_\t
W^{h^*}(\t,P_{v^i},\xx_{v^i})$ was estimated in Theorem 3.4.
Using again the
Gramm-Hadamard inequality we find

$$|V^{h^*-1}(\t)|\le L\b \e_{h^*}^n C^n \g^{2 h^*}\sum_{P_{v^1},...,P_{v^{s_v}}}
\prod_{i=1}^{s_v}[\g^{h^*({|P_{v^i}|\over 2}-2)}\g^{h^*\d_{v^i}}]$$
where $\d_{v^i}=1$ if $|P_{v^i}|=2$ and it is zero otherwise, see
Theorem 3.4,
and the theorem follows.

\vskip1.cm
\centerline{\titolo 4. The flow of the Renormalization group}
\*\numsec=4\numfor=1
{\bf 4.1} The convergence of the expansion for the effective potential
is proved by theorems 3.4,3.7 under suitable assumptions
on the running coupling constants \ie that there exists a finite $h^*$
\equ(h*) such that, given a constant $\bar\e_k$, then
$\max_{k\ge h^*}|\vec
v_{k}|\le \bar\e_k$
and $\max_{k\ge h^*}|{Z_k\over Z_{k-1}}|\le e^{c_1\bar\e_k^2}$. In this
section
we prove that it is possible to choose $|\l|,u$ so small and a proper $\nu$
so that the above conditions are indeed verified.

If $\vec a_h=\l_h,\s_h,\d_h,\t_h,\iota_h$ then
$$\vec a_{h-1}=\vec a_h+\vec\b^h(\vec a_h,\nu_h;...;\vec a_0,\nu_0)\Eq(aaa)$$
$$\nu_{h-1}=\g\nu_h+\b_\nu^h(\vec a_h,\nu_h;...;\vec a_0,\nu_0)$$
The property $\g>1$ can be used to show that
such relations are equivalent to
$$\vec a_{h-1}=\vec a_h+\vec \b^h(\vec a_h;...;\vec a_0;\nu_h)$$
$$\nu_{h-1}=\g\nu_h+\b_\nu^h(\vec v_h;...;\vec v_0;\nu_h)\Eq(bbb)$$

The proof is a trivial adaptation to this case of app.4 of
[BGPS]. For the parity of the propagator, $\b_\nu^h=\nu_h\l_h^2
\b_\nu^{1,h}+\g^h R^h_\nu$, $|\b_\nu^{1,h}|\le C$, $|R^h_\nu|\le C\bar\e_h^2$.
Moreover also from [BGPS], sec.7, it follows that
it is possible to choose $\nu\equiv\nu(\l,u)$ so that
$\nu_h=O(\g^h)$, for any choice of the running coupling constants such that
$\max_{k\ge h^*}|\vec
v_{k}|\le \bar\e_k$
and $\max_{k\ge h^*}|{Z_k\over Z_{k-1}}|\le e^{c_1\bar\e_k^2}$.
This is a version of the unstable manifold theorem.

We can write, for $h\ge h^*$
$$\l_{h-1}=\l_h+G^{1,h}_\l+G^{2,h}_{\l}+\g^h R^h_\l$$
$$\sigma_{h-1}=\sigma_h
+G_{\sigma}^{1,h}$$
$$\d_{h-1}=\d_h+G^{1,h}_\d+G^{2,h}_{\d}+\g^h R^{h}_\d\Eq(giap)$$
$$\t_{h-1}=\t_h+G^{2,h}_\t+\g^h R^{h}_\t$$
$$\iota_{h-1}=\iota_h+G^{2,h}_{\iota}+\g^h R^{h}_\iota$$
$${Z_{h-1}\over Z_h}=1+G_{z}^{1,h}+
G^{2,h}_{z}+\g^hR^{h}_z$$

where:
\vskip.5cm
1) $G^{1,h}_\l \equiv G^{1,h}_\l(\l_h,\d_h;...;\l_0,\d_0)$,
$G^{1,h}_\d\equiv G^{1,h}_\d(\l_h,\d_h;...;\l_0,\d_0)$
and

$G^{1,h}_z\equiv G^{1,h}_z(\l_h,\d_h;...;\l_0,\d_0)$ are given by series
of terms involving only the Luttinger model part of the propagator
$g^{(k)}_{L,\o}(\xx;\yy)$, $k\ge h$, see lemma 2.4.
\vskip.5cm
2) $G^{2,h}_\l,G^{2,h}_\d, G^{2,h}_z,
G^{2,h}_\t ,G^{2,h}_\iota$
depend on all the running coupling constants and are given by a series
of terms involving at least a propagator $C_{2}^{(k)}(\xx;\yy)$ or
$g^{(k)}_{\o,-\o}(\xx;\yy)$, $k\ge h$. By lemma 2.4
if $C_\a$ is a suitable constant and $\hat\e_h={\rm max}_{k\ge h}
|g_k|$:
$$|G^{2,h}_\l|, |G^{2,h}_\d|, |G^{2,h}_z|,
|G^{2,h}_\t| ,|G^{2,h}_\iota|\le C_\a \hat\e_h^2
|{\s_h\over\g^h}|\Eq(4.2)$$
\vskip.5cm
3) By a second order computation one obtains
$$G^{1,h}_\s=\s_h g_h [-\b_1+\bar G^{1,h}_\s]\quad
G^{1,h}_z=g^2_h[\b_2+\bar G^{1,h}_z]\Eq(4.3)$$
with $\b_1,\b_2$ non vanishing positive contants
and $|\bar G^{1,h}_\s|\le C_\a\hat\e_h$ and
$|\bar G^{1,h}_z|\le C_\a \hat\e_h$.
Moreover
$G^{1,h}_\l,G^{1,h}_\d$ are vanishing at the second order.
\vskip.5cm
4)$R^{h}_i$, $i=\l,z,\d,\t,\iota$
depend on all the running coupling constants and there is at least a
propagator $C_{2}^{(k)}(\xx;\yy)$, $k\ge h$.
$|R^{h}_i|\le C_\a\bar\e_h^2$.
\vskip.5cm
\0{\bf 4.2}{\rm LEMMA}: There exist positive constants $c_1,c_2,c_3,c_4$
such that, if $\l,u$ are small enough and $h\ge h^*$:
$$|\l_{h-1}-\l_{0}|<|\l|^{3/2}$$
$$\l\b_2 c_3 h< \log\left({|\sigma_{h-1}|\over
|\sigma_0|}\right)< \l\b_2 c_4 h \Eq(4.2)$$
$$-\b_1 c_1\l^2h < \log(|Z_{h-1}|)< -\b_1 c_2\l^2h\qquad
|\t_{h-1}-\t_{0}|<|\l|^{3/2}$$
$$|\iota_{h-1}-\iota_{0}|<|\l|^{3/2}\quad
|\d_{h-1}-\d_{0}|<|\l|^{3/2}$$
\vskip1cm
{\rm PROOF} We proceed by induction. Assume that the above inequalities
hold for any
$k\ge h$ and we prove them for $h-1$. Then by \equ(4.3)
$$(1-\b_2 \l_h-c_a\l^2)\le{\sigma_{h-1}\over\s_h}
\le (1-\b_2
\l_h+c_b \l^2)$$
with $c_b,c_a>0$ are suitable constants.
Then, if $\l>0$ (for fixing ideas)
there exists a $c_4<1$ such that
$$|\sigma_{h-1}|\le |\sigma_0| \g^{c_4\l\b_2(h-1)}{(1-\b_2 \l+c_b\l^2)
\over \g^{-\l
c_4\b_2}}\le
|\sigma_0| \g^{\l c_4\b_2(h-1)}$$
and a $c_3>1$ such that
$$|\sigma_{h-1}|\ge |\sigma_0|
\g^{c_3\l\b_2(h-1)}{(1-\b_2 \l-c_a\l^2)\over \g^{-\l c_3\b_2}}\ge
|\sigma_0| \g^{c_4\l\b_2(h-1)}$$
We proceed in the same way for $Z_h$. Moreover let be $\mu_h=(g_h,\d_h)$
and we write
$$G^{1,h}_\mu(\mu_h,\mu_{h+1},...,\mu_0)=G^{1,h}_\mu(\mu_h,...,\mu_h)+
\sum_{k=h+1}^0 D^{h,k}\Eq(cccc)$$
where
$$D^{h,k}=G_\mu^{1,h}(\mu_h...,\mu_h,\mu_k,\mu_{k+1},..\mu_0)-
G_\mu^{1,h}(\mu_h,...,\mu_h,\mu_h,\mu_{k+1},...,\mu_0)\Eq(zzz)$$
It holds that
$$G_\mu^{1,h}(\mu_h,...,\mu_h)=O(\g^h)\Eq(van)$$
This remarkable cancellation is due the fact that, by definition,
$G^{1,h}_\mu(\mu_h,...,\mu_h)$ is {\it identical} to the corresponding
quantity of the Luttinger model for which \equ(van) is proved
(see eq.(7.6) of [BGPS], as a consequence of the analyticity of the
Luttinger model Beta function and of some properties of the exact
solution, see [BeGM],[BM1]). On the other hand, by the estimates of
sec.3
$$|D^{h,k}|\le C \l^2 \g^{-{k-h\over 2}}$$
so that
$$|\mu_{h-1}-\mu_h|\le C \l^2\sum_{k=h}^0\g^{-{k-h\over 2}}+C\l^2
\sum_{k=h}^0{|\sigma_k|\over\g^k}+\l^2 C \sum_{k=h}^0\g^k$$
and the lemma follows noting that, as $h\ge h^*$
$$\sum_{k=h}^0{|\s_k|\over\g^k}\le |\sigma_{h^*}|\g^{-h^*}
\sum_{k=h}^0\g^{-k+h^*}{|\sigma_k|\over\sigma_{h^*}}\le K$$
if $K$ is a constant. For $\t_h,\iota_h$
we proceed in a similar way.\qed
\vskip1.cm

As a corollary of the above estimates
on the running coupling constants we find that $h^*$ is finite:
$${\log_\g |\sigma_0|\over
1-\l\b_2 c_\a}\leq h^*\leq {\log_\g |\sigma_0|+1\over 1-\l\b_2
c_\b}\Eq(h**)$$
for some positive constants $c_\a,c_\b$. From lemma 4.2 it follows that we
can choose
$\l,u$ so small so that the conditions for the validity of theorems
4.3,3.7 are indeed fulfilled; in particular one can choose
$L,\b\ge \g^{-h^*}$.
This concludes the proof of the convergence
of the expansion for the effective potential.
\vskip.5cm
\0{\bf 4.3} From the proof of the convergence of the effective potential
and the relative bounds
it is a standard matter to deduce the bounds for the expansion
of the two point Schwinger function; we refer for this to [BGPS],
sec.6. It holds that

$$S(x,y)=S^{u.v.}(\xx;\yy)+
\sum_{h=h^*}^0\sum_{\o_1,\o_2}e^{i{p_F}(\o_1x-\o_2y)}
\cdot({g^{(h)}_{\o_1,\o_2}(\xx;\yy)\over Z_h}
+{\bar S^{(h)}_{\o_1,\o_2}(\xx;\yy)\over Z_h})\Eq(S)$$
where we call $g^{(\leq h^*)}(\xx;\yy)$ simply $g^{(h^*)}(\xx;\yy)$
and the
first addend is obtained by the integration of $\psi^{(1)}$ (
and it is bounded by ${C_N\over 1+|\xx-\yy|^N}$); $\bar
S^{(h)}_{\o_1,\o_2}(\xx;\yy)$
is given by an expansion similar to the one in sec.3 and
one can check that, for $L,\b\ge\g^{-h^*}$, $|\bar
S^{(h)}_{\o_1,\o_2}|\le A\bar\e_h\g^h{C_N\over 1+(\g^h|\xx-\yy|^)N}$,
$\bar\e_h=\sup_{k>h}|\vec v_k|$, see [BGPS]..

We are now ready to prove the main theorem in 4.1.
First of all, we note that
the r.h.s.
of \equ(S) has a meaning also if we take the formal limit $i\to\io$, $\b
\to\io$, (recall that $L=L_i$), by doing the substitution
$${1\over L_i\b} \sum_{\kk\in {\cal D}_{L_i,\b}} \rightarrow
\int_{T^1\times R} {d\kk\over (2\p)^2}\; . \Eq(4.21)$$
%

The substitution \equ(4.21), applied to \equ(S), allows to define
$g^{(h)}(\xx;\yy)$ in the limit $i\to\io$, $\b\to\io$, at least for $h\le 0$.
For $h=1$, one has to be careful, since the integral over $k_0$ is not
absolutely convergent.
However it is easy to prove, by using standard well known arguments,
that the limit as $i\to\io$ and $\b\to\io$ of $g^{(1)}(\xx;\yy)$ is well
defined for $x_0\not= y_0$ and has the same discontinuity in $x_0=y_0$ of the
same limit taken on the free propagator \equ(1.3f).
The previous considerations says that, for $\l,u$ real and small enough,
there exists the limit
$$S(\xx;\yy) = \lim_{\b\to\io \atop i\to\io} S^{L_i,\b}(\xx;\yy) \Eq(4.22)$$
and that this limit is obtained by doing the substitution \equ(4.21) in all
quantities appearing in the r.h.s. of \equ(S).
\0{\bf 4.4} Let us define
$$\h_1=-{\log(Z_{h^*})\over \log |u\hat \f_{\bar m}|}
\qquad  \h_2=-1+
{\log(\sigma_{h^*})\over \log |u\hat\f_{\bar m}|}$$
which are respectively $\b_2\l^2+O(\l^3)$
and
$\b_1\l+O(\l^2)$. We call $Z_{h^*}=\hat Z$ and $\s_{h^*}=\hat u$, see (1.6).
The bounds \equ(z.1),(1.8)  can be
obtained by \equ(S). In fact if $1\le |\xx-\yy|\le\hat u^{-1}$ it holds,
for $\g^{-h_x-1}<|x-y|\leq \g^{-h_x}$,
$h_x>h^*$,
$$
\sum_{h=0}^{h^*} |{ g^{(h)}_{\o_1,\o_2}(\xx;\yy)\over Z_h}
+{\bar S^{(h)}_{\o_1,\o_2}(\xx;\yy)\over Z_h})|\le
A_1\left[\sum_{h=h^*}^{h_x-1} {\g^h\over
Z_h}+\sum_{h=h_x}^{0}{\g^h\over Z_h}{C_N\over \g^{Nh}|x-y|^N}\right]\le
A_2\g^{h_x(1-\h_3)}$$
On the other hand if $|\xx-\yy|\ge \hat u^{-1}$:
$$
\sum_{h=0}^{h^*} |{ g^{(h)}_{\o_1,\o_2}(\xx;\yy)\over Z_h}
+{\bar S^{(h)}_{\o_1,\o_2}(\xx;\yy)\over Z_h}) |   \le
{C_N\over |x-y|^N}\sum_{h=0}^{h^*}{\g^{-(N-1)h}\over Z_h}$$
$$\le A_2{\g^{h^*}\over Z_{h^*}}
{C_N\over (\g^{h^*}|x-y|)^N}
$$
if $A_1,A_2$ are positive constants.
From the above bounds and lemma 2.6 it follows that
$$\sum_{h=0}^{h^*}\sum_{\o_1,\o_2}e^{-i{p_F}(\o_1\xx-\o_2\yy)}
{g^{(h)}_{\o_1,\o_2}(\xx;\yy)\over Z_h}-\sum_{\o}e^{i{p_F}(\o(x-y))}
{1\over (L\b)}\sum_{\kk'\in {\cal D}_{L,\b}}
{f_h(\kk')\over -ik_0+\o k'}{1\over|\kk'|^{\h_3}}\le$$
$$C\sum_{h=0}^{h^*}[{\sigma_h\over Z_h}]\le C_1{ O(u,\hat u)\over
\hat Z}\log \hat u$$
so proving \equ(lutt).
\vskip.5cm
\0{\bf 4.5} Let us prove the decomposition \equ(1.21). By diagonalizing the
quadratic form ${
g^{(h)}(\xx,\yy)\over Z_{h-1}}=\sum_{\o_1,\o_2}
{e^{-i{p_F}(\o_1 x-\o_2 y)}\over Z_{h-1}}
g^{(h)}_{\o_1,\o_2}(\xx;\yy)$
it is possible to see that, from (2.33):
$${g^{(h)}(\xx;\yy)\over Z_{h-1}}=
{1\over (L\b)}\sum_{\kk'\in {\cal D}_{L,\b}} e^{-i\kk'(\xx-\yy)}
{\tilde f_h(k')\over Z_{h-1}}
[{F_{xy}(k',\sigma_{h-1})\over A+B}
+{F_{xy}(-k',-\sigma_{h-1})\over A-B}]$$
where
$$F_{xy}(k',\sigma_h)=\hat \phi(k',x,\sigma_h)\hat
\phi(k',-y,\sigma_h)$$ $$\hat
\phi(k',x,\sigma_h)={1\over\sqrt{2B}}[\sqrt{B-C}e^{i p_F\xx}
-
{e^{-i p_Fx}\over \sqrt{B-C}}]$$ and $A=-i k_0+\cos(p_F)(1-\cos k')$,
$B=\sqrt{(C^2+\s_h^2)}$ and
$C=v_0\sin k'$. We can rewrite the above integral in terms of the $k$
variable. Recall that, if $\o={\rm sign}(k)$ and $f_h(\kk')\not =0$,
then $k=\o p_F+k'$. Hence
$${g^{(h)}(\xx;\yy)\over Z_{h-1}}=\sum_{\o_1,\o_2}
{e^{i{p_F}(\o_1x-\o_2y)}\over Z_{h-1}}
g^h_{\o_1,\o_2}(\xx;\yy)=$$
$$={1\over (L\b)}\sum_{\kk\in{\cal D}_{L,\b}} {\tilde f_h(\kk)\over Z_{h-1}}
{\phi(k,x,\sigma_{h-1}))\phi(k,-y,\sigma_{h-1})
e^{ik_0(x_0-y_0)}
\over -ik_0-(\e(k,\sigma_{h-1})-p_F^2)}\Eq(iii)$$
where $\e(k,\sigma_h),\phi(k,x,\sigma_h)$ are defined in sec.(1.12).
Inserting \equ(iii) in (4.11) one obtains the decomposition
\equ(1.21). For clarity of exposition we have preferred to write such formula
avoiding the use of the scales. It is
clear that the $h$-dependence of $\s_{h-1},Z_{h-1}$ can be seen as a momentum
dependence, for the compact support properties of $\tilde f_h(\kk)$.
The functions $\hat u(\kk),\hat Z(\kk)$ are defined in the following way:
they are equal
to $\s_{h-1},Z_{h-1}$ for $\g^{h}\le |\kk'|< \g^{h+1}$ if $h\ge h^*$,
and to $\s_{h^*},Z_{h^*}$ for $|\kk'|\le \g^{h^*}$. With this definition
the difference between
(1.12) and \equ(iii) is in the second addend of the decomposition (1.11)
(remembering that, from (4.5), ${\s_{h-1}\over\s_h}=1+O(\l)$).

\vskip.5cm
\0{\bf 4.6} Finally the lower bound on the spectral gap can be obtained
showing that we can prove the boundedness on $S(x,y;k_0)$ for $k_0$
complex, for $|\Im(k_0)|\le{\sigma_{h^*}\over 2}$. The expansion
disussed before is not suitable for this, as the functions
$\hat f_h(\kk)$ are not analytic in $k_0$. However
one can repeat the above analysis using a decomposition
$\hat f_h(k)$ \ie not depending on $k_0$ (see [BGM1] app.4
in which are discussed the (obvious) modifications in the bounds changing
the cut-off functions). So in the support of $\hat f_h(k)$, for $h>h^*$
we have that, if $|\h|\le {\s_{h^*}\over 2}$
$$|\Re[A_h(k',k_0+i\h]|\ge a_0\g^{2h}+\s_h^2-\h^2\ge c_1\g^{2h}$$
as, by \equ(h*), $\g^h>\g^{h^*}\ge {4\s_{h^*}\over a_0\g}$. On the other hand
if $h\le h^*$
$$|\Re[A_h(k',k_0+i\h]|\ge {c_2\over\g^{h^*}}$$
for some constant $c_1,c_2$, so everything is essentially unchanged.
\vskip1cm
{\bf Acknowledgements} I thank G.Benfatto and G.Gallavotti for many important
discussions.
\vskip1cm
\centerline{\titolo References}
\*

\halign{\hbox to 1.2truecm {[#]\hss} &
        \vtop{\advance\hsize by -1.25 truecm \0#}\cr

A& {S. Aubry  in  Polarons and Bipolarons
in {\it Hich $T_c$ superconductors and related materials}, editors
Salye E, Alexandrov A., Liang W. }\cr

AAR& {S. Aubry, G. Abramovici, J. Raimbaut:
Chaotic polaronic and bipolaronic
states in the adiabatic Holstein model,
{\it J. Stat. Phys.} {\bf  67}, 675--780 (1992). }\cr
B& {A.D. Bryuno:
The analytic form of differential equations I,
{\it Trans. Moscow Math. Soc.} {\bf 25},
131--288 (1971); II,
{\it Trans. Moscow Math. Soc.} {\bf 26}, 199--239 (1972). }\cr
BG& {G.Benfatto, G.Gallavotti: Perturbation theory
of the Fermi surface in a quantum liquid. A general quasiparticle formalism and one
dimensional systems,
{\it J. Stat. Phys.} {\bf  59}, 541--664 (1990). }\cr

BGL& {G.Benfatto, G.Gallavotti, J. Lebowitz:
Disorder in the 1D Holstein model,
{\it Hel.Phys. Acta} (1992). }\cr

BGM1& {G.Benfatto,G.Gentile,V.Mastropietro:
{\it J. Stat. Phys.},89,655-708 (1997). }\cr

BGM2& {G.Benfatto,G.Gentile,V.Mastropietro:
preprint (1997). }\cr

BGPS& {G.Benfatto, G.Gallavotti, A.Procacci, B.Scoppola:
Beta functions and Schwinger functions for a many fermions system in one dmension,
{\it  Comm.Math.Phys.} {\bf  160}, 93--171 (1994). }\cr

BM1& {F.Bonetto,V.Mastropietro: Beta function and anomaly of the Fermi surface
for a d=1 system of interacting fermions in a periodic potential,
{\it Comm.Math.Phys.} {\bf  172}, 57-93 (1995).}\cr

BM2& {F.Bonetto,V.Mastropietro: Filled band Fermi systems
{\it Mat.Phys.Elect.Jour} {\bf 2}, 1--43 (1996 ).}\cr

BM3& {F.Bonetto,V.Mastropietro: Critical indices in a d=1 filled
band Fermi system,
{\it Phys. Rev. B} {\bf 56},3,1296-1308 (1997).}\cr
BM4& {F.Bonetto,V.Mastropietro: Critical indices for the Yukawa2 quantum field theory,
{\it Nucl.Phys.B} {\bf 497}, 541-554  (1997). }\cr

BeGM& {G.Benfatto, G.Gallavotti,V.Mastropietro:
Renormalization group and the Fermi surface in the Luttinger model,
{\it Phys. Rev. B} {\bf 45}, 5468-5480, (1992). }\cr

BLT& {J. Belissard, R. Lima, D. Testard:
A metal-insulator transition for almost Mathieu model,
{\it Comm. Math. Phys.} {\bf 88}, 207--234 (1983). }\cr
CF& {L. Chierchia, C. Falcolini:
A direct proof of a theorem by Kolmogorov in hamiltonian systems,
{\it Ann. Scuola Norm. Super. Pisa Cl. Sci.} {\bf 21},
No. 4, 541--593 (1995). }\cr

D& {H. Davenport:
{\sl The Higher Arithmetic}, Dover,
New York, 1983.}\cr
DS& {E.I. Dinaburg, Ya.G. Sinai:
On the one dimensional Schroedinger equation
with a quasiperiodic potential,
{\it Funct. Anal. and its Appl.} {\bf 9}, 279--289 (1975). }\cr
E& {L.H. Eliasson:
Floquet solutions for the one dimensional
quasi periodic Schroedinger equation,
{\it Comm. Math. Phys.} {\bf 146}, 447--482 (1992). }\cr

E1& {L.H. Eliasson:
Absolutely convergent series expansions
for quasi-periodic motions, Report 2--88, 1--31,
Department of Mathematics, University of Stockholm (1988) and
{\it Mat.Phys.Elect.Jour} {\bf 2} .}\cr

FMRS& {J. Feldman, J.Magnen, V.Rivasseau, E.Trubowitz:
An infinite volume expansion for many fermions Freen functions,
{\it Helv.Phys.Acta.} {\bf 65} 679-721,(1992)}\cr

G1& {G. Gallavotti:
Twistless KAM tori,
{\it Comm. Math. Phys.} {\bf 164}, 145--156 (1994). }\cr

G2& {G. Gallavotti: Renormalization group and ultraviolet stability
for scalar fields via renormalization group methods
{\it Rev. Mod. Phys.} {\bf 57} 471-562 (1985) }\cr
GM& {G. Gentile, V. Mastropietro:
Methods for the analysis of the Lindstedt series for KAM tori
and renormalizability in classical mechanics. A review with
some applications,
{\it Rev. Math. Phys.} {\bf 8}, 393--444 (1996). }\cr
H& {T. Holstein:
Studies of polaron motion, part 1.
The molecular-crystal model.
{\it Ann. Phys.} {\bf 8}, 325--342, (1959). }\cr
JM& {R.A. Johnson, J. Moser: The rotation number for almost periodic
potentials, {\it Commun. Math. Phys.} {\bf 84}, 403--438 (1982).}\cr
KL& {T. Kennedy, E.H. Lieb:
An itinerant electron model with crystalline or
magnetic long range order,
{\it Physica A} {\bf 138}, 320--358 (1986). }\cr
%
%

L& { Lee P.A. Nature 291,11-12,1981. }\cr
%
%
ML& {V.Mastropietro: Interacting soluble Fermi systems in one dimension,
{\it Nuovo Cim.B} {\bf 1}, 304-312, (1994). }\cr

ML& {D.Mattis, E.Lieb: Exact solution of a many fermion system
and its associated boson field,
{\it Journ.Math.Phys.} {\bf 6}, 304-312, (1965). }\cr

MP& {J. Moser, J. P\"oschel:
An extension of a result by Dinaburg and Sinai on
quasi periodic potentials,
{\it Comment. Math. Helv.} {\bf  59}, 39--85 (1984). }\cr
NO& {J.W. Negele, H. Orland:
{\sl Quantum many-particle systems}, Addison-Wesley,
New York, 1988. }\cr
PF& {L. Pastur, A. Figotin:
{\sl Spectra of random and almost periodic operators},
Springer, Berlin, 1991. }\cr
P& {R.E. Peierls:
{\sl Quantum theory of solids},
Clarendon, Oxford, 1955. }\cr
%
%
}
\*
\ciao